\documentclass[twocolumn,prd,superscriptaddress,altaffilletter,nofootinbib]{revtex4-1}
\usepackage{amsmath}
\usepackage{amssymb}
\usepackage{amsfonts}
\usepackage{comment}
\usepackage{graphicx}
\usepackage[colorlinks=true,citecolor=blue,urlcolor=blue]{hyperref}
\usepackage{acronym}
\usepackage[usenames,dvipsnames]{xcolor}
\usepackage{xspace}
\usepackage{tikz}
\usepackage{siunitx}
\usepackage{makecell}
\usepackage{multirow}
\usepackage{bm}
\usepackage{romannum}
\usepackage{xcolor}
\usepackage{orcidlink}
\usepackage[caption=false]{subfig}
\usepackage{ulem}

\usetikzlibrary{arrows,shapes,trees,decorations.pathreplacing}
\allowdisplaybreaks

\newcommand{\Msun}{M_{\odot}}
\newcommand{\e}{\mathrm{e}}
\newcommand{\iu}{\mathrm{i}}
\newcommand{\tc}{t_{\mathrm{c}}}

\newcommand{\gstlal}{\texttt{GstLAL}}
\newcommand{\bayestar}{\texttt{BAYESTAR}}
\newcommand{\bilby}{\texttt{BILBY}}
\newcommand{\dynesty}{\texttt{DYNESTY}}
\newcommand{\tf}{\texttt{TaylorF2}}
\newcommand{\phenomd}{\texttt{IMRPhenomD}}
\newcommand{\phenomp}{\texttt{IMRPhenomPv2}}
\newcommand{\phenompnrtidal}{\texttt{IMRPhenomPv2$\_$NRTidalv2}}
\newcommand{\phenomxphm}{\texttt{IMRPhenomXPHM}}
\newcommand{\NL}{N_{\mathrm{L}}}
\newcommand{\NQ}{N_{\mathrm{Q}}}
\newcommand{\flow}{f_{\mathrm{low}}}
\newcommand{\fhigh}{f_{\mathrm{high}}}
\newcommand{\ceil}[1]{\left\lceil #1 \right\rceil}                              
\newcommand{\floor}[1]{\left\lfloor #1 \right\rfloor}  

\begin{document}


\title{Rapid localization and inference on compact binary coalescences with the Advanced LIGO-Virgo-KAGRA gravitational-wave detector network}

\author{Soichiro Morisaki \orcidlink{0000-0002-8445-6747}}
\affiliation{Institute for Cosmic Ray Research, The University of Tokyo, 5-1-5 Kashiwanoha, Kashiwa, Chiba 277-8582, Japan}
\affiliation{Leonard E.\ Parker Center for Gravitation, Cosmology, and Astrophysics, University of Wisconsin-Milwaukee, Milwaukee, WI 53201, USA}
\author{Rory Smith
\orcidlink{0000-0001-8516-3324}}
\affiliation{School of Physics and Astronomy, Monash University, VIC 3800, Australia}
\affiliation{OzGrav: The ARC Centre of Excellence for Gravitational Wave Discovery, Clayton VIC 3800, Australia}
\author{Leo Tsukada \orcidlink{0000-0003-0596-5648}}
\affiliation{Department of Physics, The Pennsylvania State University, University Park, PA 16802, USA}
\affiliation{Institute for Gravitation and the Cosmos, The Pennsylvania State University, University Park, PA 16802, USA}
\author{Surabhi Sachdev \orcidlink{0000-0002-0525-2317}}
\affiliation{School of Physics, Georgia Institute of Technology, Atlanta, GW 30332, USA}
\affiliation{Leonard E.\ Parker Center for Gravitation, Cosmology, and Astrophysics, University of Wisconsin-Milwaukee, Milwaukee, WI 53201, USA}
\author{Simon Stevenson}
\affiliation{Centre for Astrophysics and Supercomputing, Swinburne University of Technology, Hawthorn, VIC 3122, Australia}
\affiliation{OzGrav: The ARC Centre of Excellence for Gravitational Wave Discovery, Clayton VIC 3800, Australia}
\author{Colm Talbot \orcidlink{0000-0003-2053-5582}}
\affiliation{LIGO Laboratory, Massachusetts Institute of Technology, 185 Albany St, Cambridge, MA 02139, USA}
\affiliation{Department of Physics and Kavli Institute for Astrophysics and Space Research, Massachusetts Institute of Technology, 77 Massachusetts Ave, Cambridge, MA 02139, USA}
\author{Aaron Zimmerman}
\affiliation{Center for Gravitational Physics, University of Texas at Austin, Austin, TX 78712, USA}

\begin{abstract}

We present a rapid parameter estimation framework for compact binary coalescence (CBC) signals observed by the LIGO-Virgo-KAGRA (LVK) detector network. The goal of our framework is to enable optimal source localization of binary neutron star (BNS) signals in low latency, as well as improve the overall scalability of full CBC parameter estimation analyses. Our framework is based on the reduced order quadrature (ROQ) technique, and resolves its shortcomings by utilizing multiple ROQ bases in a single parameter estimation run. We have also developed sets of compact ROQ bases for various waveform models, \phenomd{}, \phenomp{}, \phenompnrtidal{}, and \phenomxphm{}. We benchmark our framework with hundreds of simulated observations of BNS signals by the LIGO-Virgo detector network, and demonstrate that it provides accurate and unbiased estimates on BNS source location, with a median analysis time of $6$ minutes. The median searched area is reduced by around 30$\%$ compared to estimates produced by  \bayestar{}: from $21.8\,\mathrm{deg^2}$ to $16.6\,\mathrm{deg^2}$. Our framework also enables detailed parameter estimation taking into account gravitational-wave higher multipole moments, the tidal deformation of colliding objects, and detector calibration errors of amplitude and phase with the time scale of hours. Our rapid parameter estimation technique has been implemented in one of the LVK parameter estimation engines, \bilby, and is being employed by the automated parameter estimation analysis of the LVK alert system.

\end{abstract}

\maketitle

\acrodef{gw}[GW]{gravitational wave}
\acrodefplural{gw}[GW]{gravitational waves}
\acrodef{bh}[BH]{black hole}
\acrodef{cbc}[CBC]{compact binary coalescence}
\acrodef{bbh}[BBH]{binary black hole}
\acrodef{nsbh}[NSBH]{neutron star--black hole}
\acrodef{bns}[BNS]{binary neutron star}
\acrodef{imbh}[IMBH]{intermediate-mass black hole}
\acrodef{ligo}[LIGO]{the Laser Interferometer Gravitational-wave Observatory}
\acrodef{lvk}[LVK]{the LIGO Scientific, Virgo and KAGRA Collaboration}
\acrodef{hlv}[HLV]{LIGO Hanford--LIGO Livingston--Virgo}
\acrodef{o2}[O2]{the second observing run}
\acrodef{o3}[O3]{the third observing run}
\acrodef{o3a}[O3a]{the first half of the third observing run}
\acrodef{o4}[O4]{the fourth observing run}
\acrodef{o5}[O5]{the fifth observing run}
\acrodef{snr}[SNR]{signal-to-noise ratio}
\acrodef{csd}[CSD]{cross spectral density}
\acrodef{psd}[PSD]{one-sided power spectral density}
\acrodef{pdf}[PDF]{probability density function}
\acrodef{gwtc}[GWTC]{Gravitational Wave Transient Catalog}
\acrodef{vt}[$VT$]{sensitive space-time volume}
\acrodef{far}[FAR]{false alarm rate}
\acrodef{roq}[ROQ]{reduced order quadrature}
\acrodef{mcmc}[MCMC]{Markov chain Monte Carlo}
\acrodef{ks}[KS]{Kolmogorov-Smirnov}

\section{Introduction} \label{sec:introduction}

The first joint observation of \acp{gw} and electromagnetic waves from a \ac{bns} merger has revolutionized relativistic astrophysics \cite{LIGOScientific:2017ync}. \ac{gw} emission encoded the dynamics of the colliding objects \cite{LIGOScientific:2017vwq, LIGOScientific:2018hze} while electromagnetic emission encoded the rich physics of the subsequent short gamma-ray burst \cite{Goldstein:2017mmi, Savchenko:2017ffs, Monitor:2017mdv}, kilonova \cite{Cowperthwaite:2017dyu, Evans:2017mmy, Arcavi:2017xiz, Utsumi:2017cti}, and afterglow of the merger-remnant \cite{Alexander:2017aly, Troja:2017nqp}. These complementary observations provided information about the origins of short gamma-ray bursts and heavy elements \cite{Drout:2017ijr, Kasliwal:2017ngb, Tanvir:2017pws, Pian:2017gtc, Tanaka:2017qxj}, matter with supra-nuclear densities \cite{LIGOScientific:2018cki, LIGOScientific:2019eut}, the expansion rate of the Universe \cite{LIGOScientific:2017adf, Hotokezaka:2018dfi}, and the properties of gravity \cite{Creminelli:2017sry, Ezquiaga:2017ekz, Baker:2017hug}. More joint \ac{gw} and electromagnetic observations -- so-called \textit{multi-messenger observations} -- of \ac{cbc} events are required for more accurate and precise understanding of those topics. Increasing the number of the successful multi-messenger observations is one of the main goals of \ac{o4} of \ac{lvk} \cite{KAGRA:2013rdx, LIGOScientific:2014pky, VIRGO:2014yos, KAGRA:2020agh}, which is currently ongoing.


Rapid and accurate source localization from \ac{gw} data is key to successful multi-messenger observations of \ac{cbc} events. Additionally, as the global GW detector network improves in sensitivity and detection rates reach around one per day \cite{Petrov:2021bqm}, rapid and accurate parameter estimation on \textit{all} compact binaries ensures that data analysis scales commensurably with increasing detections. Reducing the computational cost of source-parameter estimation has been essential for making rapid and accurate parameter estimation a reality. 
Several techniques have been developed over the past years to this end. These include likelihood approximation \cite{Cornish:2010kf, Cornish:2021wxy, Leslie:2021ssu, Vinciguerra:2017ngf, Morisaki:2021ngj, Pathak:2022ktt}, parallelized algorithms \cite{Pankow:2015cra, Lange:2018pyp, Wysocki:2019grj, Wofford:2023iwz, Rose:2022axr, Smith:2019ucc, Talbot:2019okv}, machine learning approaches \cite{Green:2020hst, Dax:2021tsq, Dax:2022pxd, Williams:2021qyt, Chatterjee:2022ggk}, re-parameterizations to remove parameter degeneracy and multi-modalities \cite{Lee:2022jpn, Roulet:2022kot}, and other techniques \cite{Islam:2022afg, Wong:2023lgb, Tiwari:2023mzf}. While each of these methods has reduced the wall-time or CPU/GPU-time cost of parameter estimation (or both) to some degree, they generally require constant updating or modification to reflect progress in, e.g., developments of new model gravitational waveforms; time-sensitivity of particular observations (such as observing EM counterparts); or scalability with increasing event rate. The focus of this work is to present a flexible set of approximate methods for parameter estimation on multiple GW sources for the foreseeable future of LVK observing runs. 

Here we focus on the \ac{roq} \cite{Canizares:2014fya, Smith:2016qas} method, which accelerates parameter estimation by significantly reducing the amount of waveform evaluations -- the dominant runtime cost. The key ingredient in \ac{roq} is a re-representation of waveform models as a weighted sum over basis elements and coefficients. The latter contain the waveform's parametric dependence on the CBC's physical parameters, e.g., masses and spins. The smaller the basis size (number of basis elements), the more parameter estimation is accelerated.
Previous work \cite{Morisaki:2020oqk} has demonstrated that the basis size is drastically reduced if the ROQ basis is constructed over a targeted narrow mass-spin space, reducing the run time of parameter estimation on \ac{bns} to tens of minutes. Parameter estimation provides optimal (in the sense of minimizing the uncertainty) and unbiased sky localization of compact binaries, allowing the odds of discovering an electromagnetic counterpart to be improved if updated sky maps can be quickly disseminated to observers. Analyzing only a restricted region of the mass-spin parameter space may lead to biases in the inference if the data have support outside of the explored region. In addition, the previous work \cite{Morisaki:2020oqk} made use of a simple waveform model which does not take into account binary merger dynamics, neutron-star tidal deformability, or generic spin configurations.

In this paper we present a rapid parameter estimation framework overcoming the shortcomings of the previous approaches, which enables accurate source localization of \ac{bns} signal within minutes and greatly improves the scalability of the detailed parameter estimation analysis taking into account general binary merger dynamics. The core idea of our framework is to employ multiple \ac{roq} bases in a single parameter estimation analysis: Each basis is constructed in a targeted parameter space to gain a significant speed up, and the union of small patches in parameter space is broad enough to cover the region consistent with observed data. We also present sets of targeted \ac{roq} bases we have developed for use with our optimized framework. Some of the bases have been constructed for computationally cheap waveform models to enable rapid sky localization, and the others for the state-of-the-art waveform models taking into account gravitational-wave higher multipole moments or tidal deformation of colliding objects.

Our rapid parameter estimation technique has been implemented in one of the \ac{lvk} parameter estimation engines, \bilby{} \cite{Ashton:2018jfp, Romero-Shaw:2020owr}, and that technique as well as our newly developed \ac{roq} bases are being employed by the automated parameter estimation analysis of the \ac{lvk} \ac{o4} alert system, circulating source location estimates to follow-up observers \cite{2023GCN.33816....1L, 2023GCN.33891....1L, 2023GCN.33919....1L, 2023GCN.34087....1L}. The typical analysis time is several minutes for \ac{bns}, and hours for the other types of signal. In practice, there can be a delay making the results public due to human vetting of observed data and the inference results. This can increase the time to send out an update GCN notice/circular to an hour to several hours. However, in the future human intervention may be removed from the process so that they are circulated immediately after parameter estimation is completed.


The rest of the paper is organized as follows. In Section \ref{sec:roq}, we review the basics of \ac{roq} and describe our optimizations to the \ac{roq} method. In Section \ref{sec:bases}, we present our new \ac{roq} bases and describe how they have been constructed. In Section \ref{sec:application}, we benchmark our optimized \ac{roq} method with hundreds of simulated signals. Finally, in Section \ref{sec:conclusion}, we conclude this paper with summarizing our results. Throughout this paper, we apply the geometric unit system, $c=G=1$.

\section{Improved reduced order quadrature} \label{sec:roq}

In this section, we explain the basics of \ac{roq} and present our idea of using multiple \ac{roq} bases in a single parameter estimation run.

\subsection{Basics}

Parameter estimation of \ac{cbc} signal is typically based on Bayesian inference, where Bayesian posterior probability density function is computed via Bayes' theorem, 
\begin{equation}
p(\theta|\{d_i\}_{i=1}^{N_{\mathrm{det}}}) = \frac{\mathcal{L}(\{d_i\}_{i=1}^{N_{\mathrm{det}}}|\theta) \pi(\theta)}{\mathcal{Z}}.
\end{equation}
Here, $\{d_i\}_{i=1}^{N_{\mathrm{det}}}$ is a set of data from $N_{\mathrm{det}}$ detectors, $\theta$ is a set of parameters characterizing \ac{cbc} signal, $\mathcal{L}(\{d_i\}_{i=1}^{N_{\mathrm{det}}}|\theta)$ is likelihood function, $\pi(\theta)$ is prior probability density function, and $\mathcal{Z}$ is evidence. For \ac{cbc} parameter estimation, we typically assume that instrumental noise is stationary and Gaussian, and employ the Whittle likelihood \cite{whittle}, whose logarithm is given by
\begin{align}
\ln \mathcal{L} &= - \frac{1}{2} \sum_{i=1}^{N_\mathrm{det}} (d_i - h_i(\theta), d_i - h_i(\theta))_i + \mathrm{const.} \\
&= \sum_{i=1}^{N_\mathrm{det}} \left[ (d_i, h_i(\theta))_i - \frac{1}{2} (h_i(\theta), h_i(\theta))_i \right] + \mathrm{const.}
\end{align}
$(a, b)_i$ is the noise-weighted inner product,
\begin{equation}
(a, b)_i = \frac{4}{T} \Re \left[\sum_k \frac{a^\ast(f_k) b(f_k)}{S_i(f_k)} \right],
\end{equation}
where $T$ is data duration, $S(f)$ is the \ac{psd} of instrumental noise, and the sum is taken over evenly-spaced frequencies $\{f_k\}_k$ ranging from the low-frequency cutoff $\flow$ to the high-frequency cutoff $\fhigh$ with the frequency interval of $1 / T$.
The non-constant part of $\ln \mathcal{L}$ is referred to as log-likelihood-ratio,
\begin{equation}
\ln \Lambda = \sum_{i=1}^{N_\mathrm{det}} \left[ (d_i, h_i(\theta))_i - \frac{1}{2} (h_i(\theta), h_i(\theta))_i \right], \label{eq:log_likelihood_ratio}
\end{equation}
and is computed typically more than millions of times during the stochastic sampling of posterior.

The dominant computational cost of parameter estimation comes from the generation of waveform $\{\tilde{h}_i(f_k; \theta)\}_k$, which is required to compute the log-likelihood-ratio \eqref{eq:log_likelihood_ratio}.
The cost is proportional to the number of frequency points $K = (\fhigh - \flow) T + 1$, which is equal to the number of required waveform evaluations per waveform generation.
\ac{roq} reduces the number of required waveform evaluations by expressing the waveform and its squared-amplitude as linear functionals of ROQ bases,
\begin{eqnarray}
    \label{eq:rom}
     h_i(f_k;\theta',\,\tc=0) &\simeq& \sum_{I=1}^{\NL} h_i(F_I;\theta',\tc=0)\,B_{I}(f_k), \label{eq:rom_linear} \\
     |h_i(f_k;\theta)|^2 &\simeq& \sum_{J=1}^{\NQ} |h_i(\mathcal{F}_J;\theta)|^2\,C_{J}(f_k), \label{eq:rom_quadratic}
\end{eqnarray}
where $\tc$ is the coalescence time of signal, $\theta'$ is the set of the parameters except for $\tc$, $\{F_I\}_{I=1}^{\NL}$ and $\{\mathcal{F}_J\}_{J=1}^{\NQ}$ are known as \textit{empirical interpolation nodes}, and $\{B_{I}(f_k)\}_{I=1}^{\NL}$ and $\{C_{J}(f_k)\}_{J=1}^{\NQ}$ as linear and quadratic \ac{roq} bases\footnote{Note that the representations in Eq.~(\ref{eq:rom}) are often referred to as ``reduced order models'' or ROMs, of waveforms. Here we choose to avoid using the term ROM to minimize the amount of technical jargon, as we are primarily interested in the quantities derived from Eq.~(\ref{eq:rom}).}. Generally, the bases are defined over a sub-domain in parameter space. The sub-domain is typically smaller than the full parameter-space on which the waveform models themselves are defined.

Substituting the above expressions into the log-likelihood-ratio \eqref{eq:log_likelihood_ratio}, one arrives at the compressed \ac{roq} log-likelihood-ratio \cite{Canizares:2014fya, Smith:2016qas},
\begin{equation}
    \ln \Lambda_{\text{ROQ}} =  \sum_{i=1}^{N_\mathrm{det}} \left[ L_i(\theta) - \frac{1}{2}Q_i(\theta) \right] \,,
    \label{eq:likliehood_roq}
\end{equation}
where the functions $L_i(\theta)$ and $Q_i(\theta)$ are given by
 \begin{align} 
    L_i(\theta) &=  \Re \left[\sum_{I=1}^{N_L}h_i(F_I;\theta',\tc=0)\,\omega_{I,i}(t_c)\right]\,, \label{eq:linear} \\
    Q_i(\theta) &= \sum_{J=1}^{N_Q} |h_i(\mathcal{F}_J;\theta)|^2 \psi_{J,i}\,. \label{eq:quad}
\end{align}
The quantities $\omega_{I,i}(t_c)$ and $\psi_{J,i}$ are integration weights that depend only on the bases, data, and noise power spectral density:
\begin{align}
    \omega_{I,i}(t_c) &= \frac{4}{T} \sum_k \frac{d^{*}_i(f_k)B_I(f_k)} {S_i(f_k)} e^{-2\pi i f_k t_c}\,. \label{eq:omega} \\
    \psi_{J,i} &= \frac{4}{T} \sum_{k=1}^{K}\frac{\,C_{J}(f_k)}{S_i(f_k)}\,. \label{eq:Psi}
\end{align}
These data-dependent weights are a one-time, upfront calculation and can be efficiently computed using an inverse fast Fourier transform.
Since $\ln \Lambda_{\text{ROQ}}$ can be computed with waveform values at $\NL + \NQ$ frequency points, the number of required waveform evaluations is reduced by $K / (\NL + \NQ)$, and the analysis is expected to be accelerated by the same factor.

The ROQ bases, $\{B_{I}(f_k)\}_{I=1}^{\NL}$ and $\{C_{J}(f_k)\}_{J=1}^{\NQ}$, need to be pre-constructed and stored. For signal with long duration, which has large $K$, their file sizes can be a few tens of gigabytes or even larger. This gets a more serious issue when tens or hundreds of bases are constructed for different mass-spin sub-domains, as we do in this work. This practical issue can be resolved by utilizing the likelihood approximation technique developed in \cite{Morisaki:2021ngj}. In this approximation, the total frequency range is divided into $B$ frequency bands with a set of smooth window functions $\{w^{(b)}(f)\}_{b=1}^{B}$. They are constructed so that signal from the starting frequency of the $b$-th band has duration shorter than a certain duration value $T^{(b)}$, and their values can be chosen so that they are decreasing $T=T^{(1)} > T^{(2)} > \cdots > T^{(B)}$ thanks to the chirping nature of \ac{cbc} signal (the increase of frequency with time). The start and end frequencies of the frequency bands are determined based on the time-frequency relation of \ac{cbc} signal computed with a certain value of detector-frame chirp mass. 

Then, the frequency sum of $(d_i, h_i(\theta))_i$ is decomposed into sums over the $B$ frequency bands, and the sum over the $b$-th band is approximately computed with $h_i(f;\theta)$ at downsampled frequency points $f^{(b)}_k = k / T^{(b)}$. Similarly, $(h_i(\theta), h_i(\theta))_i$ is decomposed, and the sum over the $b$-th band is approximately computed with $|h_i(f;\theta)|^2$ at $\hat{f}^{(b)}_k = k / \hat{T}^{(b)},~\hat{T}^{(b)}=\min[2 T^{(b)}, T]$. Since $1 / T^{(b)} \gg 1 / T$ for large $b$, this \textit{multi-band} approximation significantly reduces the number of required waveform values at high frequency. Substituting Eqs. (\ref{eq:rom_linear}) and (\ref{eq:rom_quadratic}) into the multi-band forms of the inner products, they are approximated by \ac{roq} bases at the downsampled frequency points ($B_{I}(f)$ at $\{\{f^{(b)}\}_k\}_{b=1}^{B}$ and $C_{J}(f)$ at $\{\{\hat{f}^{(b)}\}_k\}_{b=1}^{B}$). The exact forms of the multi-banded \ac{roq} inner products are given in Appendix \ref{sec:multiband}. Hence we need to store only the multi-banded basis components. For typical \ac{bns} signal in the \ac{lvk} frequency range, the original number of frequency points $K \sim 10^6$ is reduced to $\sim 10^4$ with the multi-band approximation \cite{Morisaki:2021ngj}, and hence the file size is reduced by a factor of $\sim 100$. 

\subsection{Using multiple ROQ bases for arbitrary mass-spin priors}

In previous parameter estimation analyses using \ac{roq} likelihoods, the explored parameter range has typically been set by the width of a single \ac{roq} mass-spin partition. Hence the width of \ac{roq} bases have been designed to be wide enough so that posterior distributions of a signals' mass and spin will be comfortably contained within them. This introduces a tradeoff between efficiency and accuracy: \ac{roq} basis constructed over a narrow mass-spin space has a smaller basis size, but restricts the explored parameter space.

Besides this tradeoff, there are a number of drawbacks to this approach. Analysts often chose priors based on astrophysical considerations: these might be wider than those offered by individual ROQ bases; a detection trigger might have masses which fall near the boundary of an \ac{roq} basis; diagnostic checks (such as $p$-$p$ tests) may require a consistent broad prior; and catalogues of gravitational-wave events might want to impose consistent priors for particular classes of events.

To overcome these issues, it is straightforward to define an \ac{roq} likelihood over a parameter domain larger than the small sub-domains of the individual \ac{roq} basis. One simply builds multiple \textit{sets} of \ac{roq} weights constructed from bases covering different parameter sub-domains. The likelihood over the full domain is then just the union of the likelihoods on the individual sub-domains:
\begin{widetext}
\begin{equation}   
 \mathcal{L}_{\text{ROQ}}(d|\theta, \text{all bases}) = 
\begin{cases}
    \mathcal{L}_{\text{ROQ}}(d|\theta,\text{particular basis}),& \text{for } \theta\,\text{in domain of a particular basis}\\
    0,              & \text{otherwise}
\end{cases}
\end{equation}
\end{widetext}
Thus, when a parameter sample $\theta$ is drawn, one simply computes the likelihood function using the pre-computed \ac{roq} weights associated with the basis set whose domain contains $\theta$.
This likelihood construction from multiple \ac{roq} bases has been implemented in \texttt{ROQGravitationalWaveTransient} of \bilby{}.
In the current implementation, \ac{roq} bases used in a single run are assumed to have the same values of $\flow$, $\fhigh$, and $1/T$, while it can in principle be generalized. 

\section{Construction of reduced order quadrature bases} \label{sec:bases}

Based on the idea of using multiple \ac{roq} bases, we have constructed sets of \ac{roq} bases constructed over targeted mass-spin sub-domains. In this section, we present those bases as well as explaining how they have been constructed. In this work, we consider dividing the parameter space based on a single parameter, detector-frame chirp mass\footnote{sometimes referred to as red-shifted chirp mass} $\mathcal{M} \equiv (m_1 m_2)^{3/5} / (m_1 + m_2)^{1/5}$, where $m_1$ and $m_2$ are detector-frame component masses satisfying $m_1 \geq m_2$. This parameter is known to predominantly determine the gravitational waveforms' phase and amplitude evolution, and groups waveforms with similar morphologies. Using more sophisticated parameters such as the linear combinations of \ac{gw} phase coefficients introduced by \cite{Morisaki:2020oqk} may reduce the basis sizes further, but we leave exploration in that direction to future work.   

\subsection{Waveforms and parameter ranges}

\begin{table*}
\centering
\begin{tabular}{c | c | c | c | c}

Mass range & Waveform model & Spin range & Tides range & Sub-domain width \\

\hline

\multirow{3}{*}{\makecell{BNS\\($0.6 M_\odot \leq \mathcal{M} \leq 4.0 M_\odot,~~1/8 \leq q \leq 1$)}} & \phenomd & $a \leq 0.05$ & -- & $\Delta \mathcal{M}^{-5/3}=0.01~(M^{-5/3}_\odot)$ \\
\cline{2-5}
 & \phenomp & \multirow{2}{*}{$a \leq 0.99$} & -- & \multirow{2}{*}{$\Delta \mathcal{M}^{-5/3}=0.1~(M^{-5/3}_\odot)$} \\
\cline{2-2} \cline{4-4}
 & \phenompnrtidal & & $\Lambda \leq 5000$ & \\
\hline
\multirow{2}{*}{\makecell{Intermediate\\($1.4M_{\odot} \leq \mathcal{M} \leq 21M_{\odot},~~1/20 \leq q \leq 1$)}} & \multirow{2}{*}{\makecell{\phenomp}} & \multirow{2}{*}{\makecell{$a \leq 0.99$}} & \multirow{2}{*}{--} & \multirow{2}{*}{\makecell{$\Delta \mathcal{M}^{-5/3}=0.01~(M^{-5/3}_\odot)$}} \\
 & & & & \\
\hline
\multirow{2}{*}{\makecell{BBH\\($10.02 M_\odot \leq \mathcal{M} \leq 200 M_\odot,~~1/20 \leq q \leq 1$)}} & \multirow{2}{*}{\makecell{\phenomxphm}} & \multirow{2}{*}{\makecell{$a \leq 0.99$}} & \multirow{2}{*}{\makecell{--}} & \multirow{2}{*}{\makecell{See Section \ref{sec:subdomain}}} \\
 & & & & \\

\end{tabular}
\caption{Waveform models and parameter ranges for which \ac{roq} bases have been constructed, and widths of chirp-mass sub-domains.}
\label{tab:bases}
\end{table*}

We consider three different mass ranges: \ac{bns}, \ac{bbh}, and  intermediate regions. For each region, we have constructed \ac{roq} bases for different waveform models and parameter ranges. All of our bases have been constructed to approximate waveform values from $\flow=20\,\mathrm{Hz}$, the default low-frequency cutoff used in \ac{lvk} parameter estimation analyses \cite{LIGOScientific:2021djp, LIGOScientific:2021usb}. They are summarized in Tab. \ref{tab:bases}.

\subsubsection{\ac{bns}}

Astrophysical \ac{bns} masses plausibly span a range between around $1\,M_{\odot}$ to $2\,M_{\odot}$, with mass ratios $q \equiv m_2 / m_1$ roughly in the range $0.5 \leq q \leq 1$ \cite{Tauris:2017omb}. The upper and lower limits are uncertain due to a limited number of galactic and extra-galactic observations, together with incomplete models of binary neutron star astrophysics. The dimensionless spin magnitudes of colliding neutron stars, $a_1$ and $a_2$, can be up to $\sim 0.7$ assuming a plausible equation of state of matter with supra-nuclear densities \cite{Lo:2010bj}, while \ac{bns} systems that have been found by electromagnetic observations and will merge within a Hubble time will have spins of 0.04 at largest when they merge \cite{Burgay:2003jj}. Tidal deformation of the stars also affects the gravitational waveform \cite{Flanagan:2007ix}, and this effect is characterized by dimensionless tidal deformability parameters, $\Lambda_1$ and $\Lambda_2$. Measurements of those values provide constraints on the uncertain equation of state of matter with supra-nuclear densities \cite{LIGOScientific:2018cki, LIGOScientific:2019eut} and are of particular interest for nuclear physics.

\begin{figure}[t]
	\centering
    \includegraphics[width=0.98\linewidth]{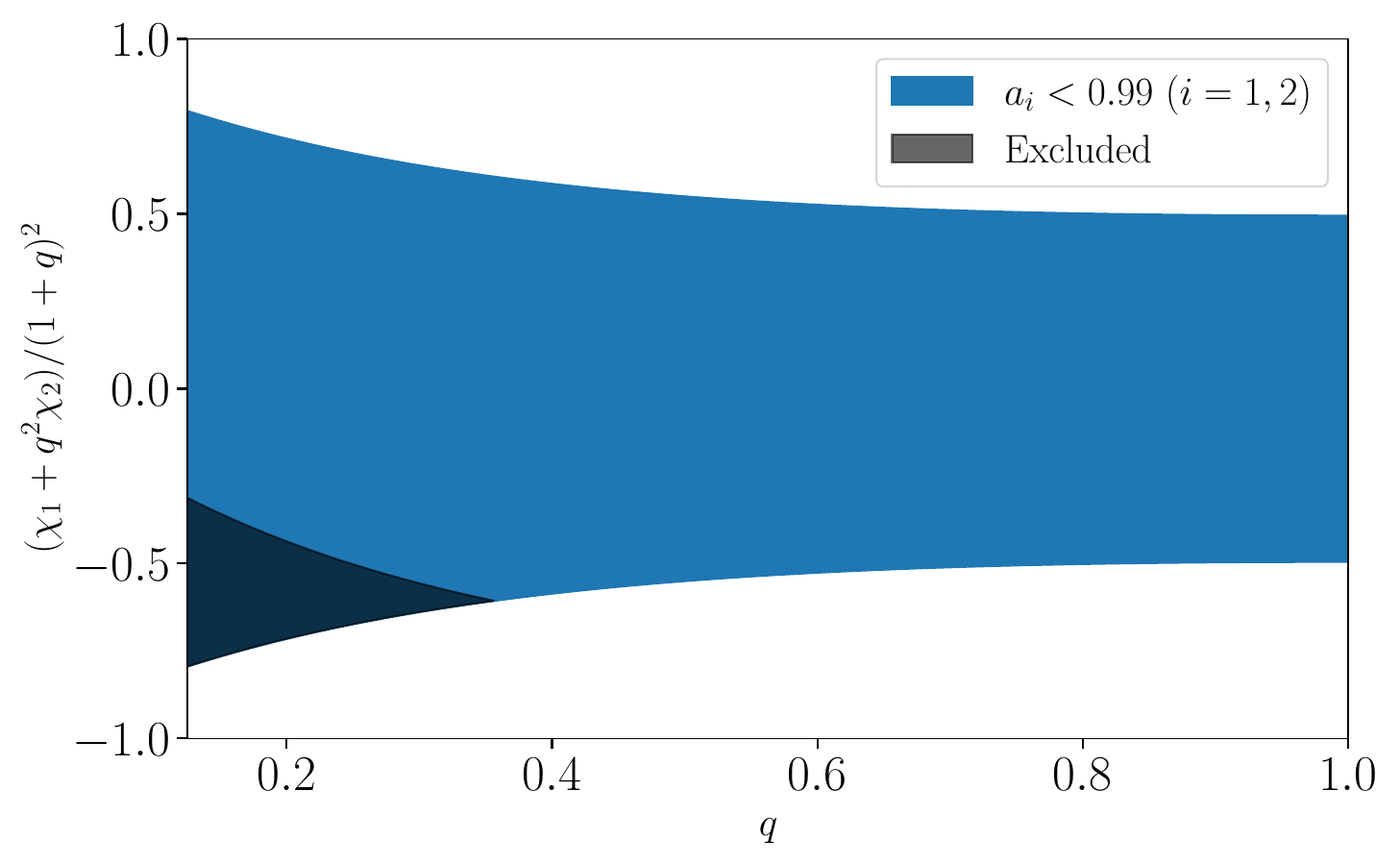}
	\caption{Excluded region of mass ratio $q$ and projected total spin $\chi=(\chi_1 + q^2 \chi_2) / (1 + q)^2$ in the basis construction of \phenomp{} and \phenompnrtidal{}, where $\chi_1$ and $\chi_2$ are dimensionless spins projected onto the orbital angular momentum. The gray region is the excluded region and the blue region is the region allowed by the spin limit $a_1,a_2 \leq 0.99$. The excluded region is determined by Eq. (\ref{eq:pv2_exclude}), and the gray region presented here is calculated with $\mathcal{M}=1M_{\odot}$ and $\fhigh=4096\,\mathrm{Hz}$, while it very weakly depends on $\mathcal{M}$ within the range we consider, $0.6 M_\odot \leq \mathcal{M} \leq 4.0 M_\odot$.}  \label{fig:exclude}
\end{figure}

Our \ac{bns} \ac{roq} bases span detector-frame chirp masses in the range $0.6 M_\odot \leq \mathcal{M} \leq 4 M_\odot$ and mass ratios in the range $1/8 \leq q \leq 1$. While this mass range extends far beyond what is plausible for astrophysical \ac{bns}{s}, our motivation is to provide an ``insurance buffer'' to accommodate unexpected sources, unusually broad posterior densities, etc. Over the mass range, we have constructed \ac{roq} bases for three different sets of waveform models and parameter ranges: \phenomd{} \cite{Husa:2015iqa, Khan:2015jqa} for the low-spin range $0 \leq a_1,a_2 \leq 0.05$, \phenomp{} \cite{Hannam:2013oca} for the high-spin range $0 \leq a_1,a_2\leq0.99$, and \phenompnrtidal{} \cite{Dietrich:2019kaq} for the high-spin range and the broad tidal deformability range $0\leq\Lambda_1,\Lambda_2\leq5000$. \phenomp{} and \phenompnrtidal{} have cusps in waveform at a certain mass-spin space, where $q \lesssim 0.4$ and spins are anti-aligned with the orbital angular momentum. As discussed in \cite{Smith:2016qas}, those cusps make it practically impossible to obtain converged \ac{roq} bases. Since the waveform models are not valid in that region anyway, we exclude that parameter space for basis construction. The excluded mass ratio--spin region for $\mathcal{M}=1M_\odot$ is shown in gray in Figure \ref{fig:exclude}, where $\chi_1$ and $\chi_2$ are spin components projected onto the orbital angular momentum, and the region is almost unchanged for a different $\mathcal{M}$ value within the range. More details about the waveform cusps will be explained in Appendix \ref{sec:pv2_cusps}.

The sole purpose of the low-spin \phenomd{} \ac{roq} bases is to provide rapid sky location for \ac{bns} candidates. \phenomd{} is valid only for simple spin configurations where spins are aligned with the orbital angular momentum. While this waveform restriction and the low-spin assumption does not allow us to explore a broader, more agnostic spin space, parameter estimation using those bases is extremely quick, enabling us to provide estimated sky location with the time scale of minutes as demonstrated in Sec. \ref{sec:application}. On the other hand, \phenomp{} is valid for general spin configurations and \phenompnrtidal{} also takes into account tidal deformation of colliding objects. Those two bases are useful for more detailed follow-up analysis of \ac{bns} candidates. 

\subsubsection{\ac{bbh}}

For the \ac{bbh} mass range, we have constructed \ac{roq} bases for \phenomxphm{} \cite{Pratten:2020ceb}, which is valid for general spin configurations and takes into account \ac{gw} higher multipole moments. Astrophysical \ac{bbh} masses observable by \ac{lvk} detectors span a range between around $2M_{\odot}$ to $400M_{\odot}$. The mass ratio distribution is subject to large uncertainties, however the \phenomxphm{} model itself accurately describes binaries with mass ratios in the range $1/20 \lesssim q \leq 1$. Current constraints on the spin magnitude and orientations of astrophysical \ac{bbh}{s} allow any magnitude up to $1$ and any possible orientation \cite{KAGRA:2021duu}. We allow our bases to span the spin magnitude range $0 \leq a_1,a_2\leq0.99$ and all possible spin directions.

Currently, the mass range of our bases spans detector-frame chirp masses in the range $10.02M_\odot \leq \mathcal{M} \leq 200M_{\odot}$ and mass ratios in the range $1/20 \leq q \leq 1$. The mass range does not extend to the lowest mass regions of the binary black hole space, which is due to current technical limitations in the design and construction of \ac{roq} bases. Specifically, building comprehensive training sets for long-duration, low mass binary black hole waveforms is more challenging than for binary neutron stars because of the presence of higher multipole moments in the signals. This requires significantly larger training sets, and hence memory, than is currently feasible. Constructing low-mass \phenomxphm{} \ac{roq} bases will be the subject of future work. 

\subsubsection{Intermediate}

To fill in the gap between the \ac{bns} and \ac{bbh} mass ranges, we have constructed \ac{roq} bases of \phenomp{} for detector-frame chirp masses in the range $1.4M_\odot \leq \mathcal{M} \leq 21M_{\odot}$. Our bases span mass ratios in the range $1/20 \leq q \leq 1$, which is broad enough to include any \ac{nsbh} binaries that may lead to electromagnetic counterparts \cite{Foucart:2020ats}, and spins in the range $0 \leq a_1,a_2 \leq 0.99$. As with the \ac{bns} bases, we exclude the mass-spin space where the waveform has cusps. These bases are being used for automated \ac{lvk} parameter estimation of event candidates which do not fall in the \ac{bns} or \ac{bbh} mass region.

\subsection{Mass-frequency partitions and multi-banding} \label{sec:mass_partitions}

Following \cite{Smith:2016qas}, we divide each mass region into several overlapping chirp mass partitions with different frequency resolution $1 / T$ and high-frequency cutoffs $\fhigh$. The chirp mass range and $T$ of each partition are determined so that the \ac{roq} bases in that partition can accurately model any waveform whose duration falls between $2^{n-1}\,\mathrm{s}$ and $2^{n}\,\mathrm{s}$, where $n$ is an integer. More mathematically, given a map from waveform duration $\tau$ to $\mathcal{M}$, $\mathcal{M}(\tau)$, the chirp mass range is determined by
\begin{equation}
\mathcal{M}\left((2^n - 2.1)\,\mathrm{s}\right) \leq \mathcal{M} \leq 1.2 \mathcal{M}\left((2^{n-1} - 2.1)\,\mathrm{s}\right), \label{eq:mc_partition}
\end{equation}
and $T = 2^n\,\mathrm{s}$. The time offset of $2.1\,\mathrm{s}$ is to accommodate the time between the coalescence time $\tc$ and the end time of analyzed data $t_{\mathrm{e}}$. Let $t_{\mathrm{trig}}$ be trigger time reported at signal detection, we typically have $2\,\mathrm{s}$ safety margin between $t_{\mathrm{trig}}$ and $t_{\mathrm{e}}$, and the standard prior of $\tc$ is uniform distribution in $t_{\mathrm{trig}} - 0.1\,\mathrm{s} \leq \tc \leq t_{\mathrm{trig}} + 0.1\,\mathrm{s}$ \cite{Veitch:2014wba}. Hence $(2 + 0.1)\,\mathrm{s}$ is the maximum time difference between $\tc$ and $t_{\mathrm{e}}$ in the standard parameter estimation, ignoring the light-traveling time between the geocenter and a detector.

For the \ac{bns} and intermediate mass regions, we employ the leading-order $\tau$-$\mathcal{M}$ relation,
\begin{equation}
\mathcal{M}(\tau) = \left(\frac{5}{256 \pi \flow \tau} \right)^{\frac{3}{5}} \frac{1}{\pi \flow}, \label{eq:tau_Mc_relation}
\end{equation}
with $\flow = 20\,\mathrm{Hz}$. With this algorithm, the \ac{bns} mass region is divided into 4 chirp-mass partitions, $0.6M_\odot \leq \mathcal{M} \leq 1.1M_\odot$, $0.92M_\odot \leq \mathcal{M} \leq 1.7M_\odot$, $1.4 M_\odot \leq \mathcal{M}  \leq 2.6 M_\odot$, and $2.1 M_\odot \leq \mathcal{M} \leq 4.0 M_\odot$, with $T=512\,\mathrm{s}$, $256\,\mathrm{s}$, $128\,\mathrm{s}$, and $64\,\mathrm{s}$ respectively. For the low-spin \phenomd{} bases, $\fhigh$ is $1024\,\mathrm{Hz}$ for all the partitions, which is high enough not to degrade estimates on source locations. For the \phenomp{} and \phenompnrtidal{} bases, $\fhigh$ is $4096\,\mathrm{Hz}$ for the first 3 partitions and $2048\,\mathrm{Hz}$ for the last partition to incorporate all the information on binary merger. The intermediate mass region is divided into 5 partitions, $1.4 M_\odot \leq \mathcal{M}  \leq 2.6 M_\odot$, $2.1 M_\odot \leq \mathcal{M} \leq 4.0 M_\odot$, $3.3 M_\odot \leq \mathcal{M} \leq 6.3 M_\odot$, $5.2 M_\odot \leq \mathcal{M}  \leq 11 M_\odot$, and $8.7 M_\odot \leq \mathcal{M}  \leq 21 M_\odot$ with $T=128\,\mathrm{s}$, $64\,\mathrm{s}$, $32\,\mathrm{s}$, $16\,\mathrm{s}$, and $8\,\mathrm{s}$ respectively. $\fhigh$ is $1024\,\mathrm{Hz}$ for all the partitions. 

\phenomxphm{} contains \ac{gw} higher multipole moments and their frequency-time relation is different from that of the dominant quadrupole moment. However, the duration of the $(l, |m|)$ modes can approximately be calculated by the same formula (\ref{eq:tau_Mc_relation}) with the frequency scaling $\flow \rightarrow (2 / |m|) \flow$. To obtain the most conservative value of $T$, we assume $|m|=4$, the highest $|m|$ of \phenomxphm{} leading to the longest duration, and employ the relation (\ref{eq:tau_Mc_relation}) with $\flow=10\,\mathrm{Hz}$. With this relation, the \ac{bbh} region is divided into 4 partitions, $10.02 M_\odot \leq \mathcal{M}  \leq 19.05 M_\odot$, $17.32 M_\odot \leq \mathcal{M} \leq 31.85 M_\odot$, $26.54 M_\odot \leq \mathcal{M} \leq 62.86 M_\odot$, and $52.38 M_\odot \leq \mathcal{M} \leq 200.0 M_\odot$ with $T=32\,\mathrm{s}$, $16\,\mathrm{s}$, $8\,\mathrm{s}$, and $4\,\mathrm{s}$ respectively.

To reduce the total file size of \ac{roq} bases with large $T$, we downsample all the \ac{bns} bases, and the \ac{nsbh} bases with $T=128\,\mathrm{s}$, $64\,\mathrm{s}$, and $32\,\mathrm{s}$ using the multi-band approximation. The frequency bands of each partition are determined based on the time-frequency relation calculated with the chirp mass value $0.95 \mathcal{M}_{\mathrm{min}}$, where $\mathcal{M}_{\mathrm{min}}$ is the minimum chirp mass of the partition and $0.95$ is a safety factor. The duration of the band decreases at the rate of $1/2$, $T^{(b)}=T/2^{b-1}$.

\subsection{Chirp mass sub-domains} \label{sec:subdomain}

Each partition is further divided into narrow $\mathcal{M}$ sub-domains to reduce the basis sizes. Each partition of the \ac{bns} and intermediate mass regions is divided equally in $\mathcal{M}^{-5/3}$, the leading order mass combination entering into \ac{gw} phasing. For the \phenomp{} and \phenompnrtidal{} bases of the \ac{bns} mass region, each partition is divided into sub-domains with the width of $\Delta \left(\mathcal{M}^{-5/3}\right)=0.1M^{-5/3}_\odot$, and one set of linear and quadratic bases have been constructed over each sub-domain. It results in $15$, $8$, $4$, and $2$ linear and quadratic bases for the partitions of $T=512\,\mathrm{s}$, $256\,\mathrm{s}$, $128\,\mathrm{s}$, and $64\,\mathrm{s}$ respectively. For the low-spin \phenomd{} bases, each partition is divided into sub-domains with the width of $\Delta \left(\mathcal{M}^{-5/3}\right)=0.01M^{-5/3}_\odot$, narrower sub-domains to obtain compact bases for low-latency source localization. We have confirmed that reducing the chirp-mass width further reduces the basis sizes only by a few tens of percents. Only the linear basis is constructed over each sub-domain. Conversely, since the quadratic basis of low-spin \phenomd{} does not significantly depend on the width of a sub-domain and its basis size is much smaller than the sizes of the linear bases, the quadratic bases have been constructed over the whole mass partitions. It results in $149$, $74$, $37$, and $20$ linear bases for the partitions of $T=512\,\mathrm{s}$, $256\,\mathrm{s}$, $128\,\mathrm{s}$, and $64\,\mathrm{s}$ respectively, and one quadratic basis per partition.

For the \phenomp{} bases of the intermediate mass region, each partition is divided into sub-domains with the width of $\Delta \left(\mathcal{M}^{-5/3}\right)=0.01M^{-5/3}_\odot$, and one set of linear and quadratic bases have been constructed over each sub-domain. It results in 37, 20, 10, 5, and 3 linear and quadratic bases for the partitions of $T=128\,\mathrm{s}$, $64\,\mathrm{s}$, $32\,\mathrm{s}$, $16\,\mathrm{s}$, and $8\,\mathrm{s}$ respectively.

For \phenomxphm{}, we take a somewhat more \textit{ad hoc} approach than for the other waveform models used in this paper. Studies to optimize the widths of sub-domains are still ongoing, and we present the results for the bases which are currently being used in the \ac{lvk} automated parameter estimation analysis. The 4s, 8s and 16s mass spaces are split into 11 equally sized sub-domains, chosen to manage memory and computational resources. The 32s mass space is split into 24 equal sub-domains. We have found empirically that it yields comparably sized bases sets, though we note that they are not optimal in the sense that further reduction in size will likely yield more compact sets. This is the topic of future work.

\subsection{Basis construction}

For constructing \phenomxphm{} bases, we employ the same strategy and codebase as used in \cite{Smith:2016qas}. The basis construction is a combination of the \textit{greedy algorithm} \cite{Field:2011mf} and the \textit{empirical interpolation method}. First, the greedy algorithm is run on a set of randomly drawn \ac{cbc} waveforms, and $N$ reduced basis vectors are obtained, whose span can approximate any waveform in the set within a specified accuracy. The set of waveforms used for constructing the reduced basis is referred to as \textit{training set}. Next, the empirical interpolation method uses the $N$ reduced basis vectors to construct $N$ empirical interpolation nodes and \ac{roq} basis. The interpolant is then \textit{validated} by computing its representation errors for waveforms outside the training set. If there are waveforms whose errors exceed an error tolerance, these waveforms are added to the training set, and the whole process is repeated. In the \phenomxphm{} model, waveform morphology is determined by the $10$ parameters, two masses, two spin vectors, the inclination angle of the orbital plane, and coalescence phase. Hence the training and validation sets consist of waveforms with random realizations of those $10$ parameter values.

For the \ac{bns} and \ac{nsbh} bases, we employ a slightly different strategy. We skip the empirical interpolation method in the iterative loop, and validate reduced basis vectors by computing their projection errors for waveforms outside the training set. Once reduced basis vectors passing the validation test are obtained, they are mapped to empirical nodes and \ac{roq} basis with the empirical interpolation method.

Amplitude and phase deviations due to detector calibration errors \cite{Vitale:2011wu} are also taken into account for the \ac{bns} and \ac{nsbh} bases. Those deviation factors are randomly realized and multiplied by a certain fraction of waveforms in the training and validation sets. They are calculated via spline interpolation of their values at 10 nodal frequency points distributed log-uniformly. Their values at the nodes are drawn from uniform distribution from $-20\%$ to $20\%$ for amplitude and $-15\,\mathrm{deg}$ to $15\,\mathrm{deg}$ for phase. While they are not considered in the basis construction for \phenomxphm{}, the bases are shown to be accurate to represent waveforms incorporating those deviations in Section \ref{sec:likelihood_errors}.

For \phenomd{}, \ac{gw} polarizations, $h_+(f)$ and $h_{\times}(f)$, are linearly dependent. Hence the reduced bases need to be constructed only for $h_+(f)$ and $\left|h_+(f)\right|^2$. Its waveform morphology is parameterized by the 4 parameters $(m_1, m_2, \chi_1, \chi_2)$. The initial training set consists of $3 \times 5^4 = 1875$ waveforms, where one third of them are waveforms on $5^4$ grid points of $\mathcal{M}$-$q$-$\chi_1$-$\chi_2$ space with vanishing calibration errors, and two thirds of them are waveforms on the same mass-spin grid points with random calibration errors. The reduced basis is validated against $3\times10^4$ waveforms, where one third of them are with vanishing calibration errors, and two thirds of them are with random calibration errors. 

As explained in a previous sub-section, all the $\phenomd{}$ bases are multi-banded. Rather than constructing \ac{roq} basis for fully sampled waveforms and downsampling it, we downsample the waveforms in the training and validation sets and directly construct multi-banded \ac{roq} basis. To reflect the integration weights in the multi-band inner products, (\ref{eq:multiband_linear}) and (\ref{eq:multiband_quadratic}), we also multiply $\sqrt{w^{(b)}(f^{(b)}_k) / T^{(b)}}$ and $\sqrt{w^{(b)}(\hat{f}^{(b)}_k) / \hat{T}^{(b)}}$ by each component of waveforms for linear and quadratic basis construction respectively. The waveforms are normalized in the following ways for linear and quadratic basis construction respectively, 
\begin{align}
&\sum_{b=1}^{B} \frac{1}{T^{(b)}} \sum_k w^{(b)}(f^{(b)}_k) \left|h_+(f^{(b)}_k)\right|^2 = 1, \\
&\sum_{b=1}^{B} \frac{1}{\hat{T}^{(b)}} \sum_k w^{(b)}(\hat{f}^{(b)}_k) \left|h_+(\hat{f}^{(b)}_k)\right|^2 = 1.
\end{align}
The error tolerances for projection errors are $10^{-10}$ for all the linear bases and $6.4 \times 10^{-15}$, $9.8 \times 10^{-15}$, $1.4 \times 10^{-14}$, and $1.7 \times 10^{-14}$ for quadratic bases with $T=512\,\mathrm{s}$, $256\,\mathrm{s}$, $128\,\mathrm{s}$, and $64\,\mathrm{s}$ respectively. Those tolerance values are empirically determined so that the relative log-likelihood-ratio errors measured in the next sub-section are $\lesssim 10^{-4}$.

For \phenomp{} or \phenompnrtidal{}, $h_+(f)$ and $h_{\times}(f)$ are linear combinations of the following 5 base waveforms,
\begin{equation}
\begin{aligned}
&l_m(f) = \e^{\iu \left(m \alpha(f) - 2 \epsilon(f)\right)} d^2_{2, m}\left(-\beta(f)\right) h_{\mathrm{D}} (f), \\
&(m=-2,-1, 0, 1, 2), \label{eq:pv2_l}
\end{aligned}
\end{equation}
where $\alpha(f)$, $\beta(f)$, and $\epsilon(f)$, are Euler angles to parametrize the rotation from an inertial frame whose $z$ axis is aligned with the total angular momentum to a co-precessing frame whose $z$ axis is aligned with the orbital angular momentum, $d^l_{m',m}(\beta)$ is the component of the Wigner matrix, and $h_{\mathrm{D}} (f)$ is the waveform in the co-precessing frame computed with the \phenomd{} model. 
The basis vectors are constructed to approximate those 5 base waveforms instead of the original polarizations.
The base waveforms are parameterized by the 5 parameters $\left(m_1,m_2,\chi_1,\chi_2,\chi_{\mathrm{p}}\right)$ for \phenomp{} and the 7 parameters $\left(m_1,m_2,\chi_1,\chi_2,\chi_{\mathrm{p}}, \Lambda_1, \Lambda_2\right)$ for \phenompnrtidal{}, where $\chi_{\mathrm{p}}$ is the effective precessing spin parameter \cite{Schmidt:2014iyl}.
They do not include the angle between the line-of-sight and the total angular momentum, $\theta_J$, and the initial phase of $\alpha(f)$, $\alpha_0$, while those 2 parameters need to be taken into account when basis vectors are constructed for the original polarizations \cite{Smith:2016qas}.
Hence building the basis vectors for the base waveforms reduces the number of parameters by 2.

On the other hand, $\left|F_+ h_+ + F_{\times} h_{\times} \right|^2$, where $F_+$ and $F_{\times}$ are detector beam pattern functions, is the linear combination of the following base waveforms,
\begin{align}
&\begin{aligned}
&q^{\mathrm{cos}}_{m,m'}(f) = \\
&~~~~\bigg[d^2_{2,m}\left(-\beta(f)\right) d^2_{2,m'}\left(-\beta(f)\right) + \\
&~~~~~(-1)^{m + m'} d^2_{2,-m}\left(-\beta(f)\right) d^2_{2,-m'}\left(-\beta(f)\right) \bigg] \times \\
&~~~~\cos \left[ (m - m') \alpha(f) \right] \left| h_{\mathrm{D}} (f) \right|^2, \\
\end{aligned}  \label{eq:pv2_q_cos} \\
&\begin{aligned}
&q^{\mathrm{sin}}_{m,m'}(f) = \\
&~~~~\bigg[d^2_{2,m}\left(-\beta(f)\right) d^2_{2,m'}\left(-\beta(f)\right) + \\
&~~~~~(-1)^{m + m'} d^2_{2,-m}\left(-\beta(f)\right) d^2_{2,-m'}\left(-\beta(f)\right) \bigg] \times \\
&~~~~\sin \left[ (m - m') \alpha(f) \right] \left| h_{\mathrm{D}} (f) \right|^2,
\end{aligned} \label{eq:pv2_q_sin} \\
&(m,m'=-2,-1, 0, 1, 2). \nonumber
\end{align}
Since $q^{\mathrm{cos}}_{m,m'}=q^{\mathrm{cos}}_{m',m}=q^{\mathrm{cos}}_{-m,-m'}$ and $q^{\mathrm{sin}}_{m,m'}=-q^{\mathrm{sin}}_{m',m}=-q^{\mathrm{sin}}_{-m,-m'}$, only 15 of them are linearly independent. Thus basis vectors are constructed for the 15 base waveforms. Those base waveforms are parametrized by the same parameters as those parametrizing $\{l_m(f)\}_{m=-2}^{2}$.

For \phenomp{} or \phenompnrtidal{} basis, we start with a training set of $\mathcal{O}(10^4)$ of waveforms. They are generated with $\mathcal{O}(10^3)$ random source parameters, for each of which there are $5$ base waveforms for linear basis and $15$ for quadratic basis. Validation is carried out with $10^5$--$10^6$ random source parameters for  \phenomp{} and $10^6$--$10^7$ random source parameters for  \phenompnrtidal{}, where half of them incorporate random detector calibration errors. The base waveforms are normalized so that their aligned-spin limits have norm of unity,
\begin{align}
&\sum_{b=1}^{B} \frac{1}{T^{(b)}} \sum_k w^{(b)}(f^{(b)}_k) \left|h_{\mathrm{D}}(f^{(b)}_k)\right|^2 = 1, \\
&\sum_{b=1}^{B} \frac{1}{\hat{T}^{(b)}} \sum_k w^{(b)}(\hat{f}^{(b)}_k) \left|h_{\mathrm{D}}(\hat{f}^{(b)}_k)\right|^2 = 1.
\end{align}
For the \ac{bns} \phenomp{} bases, the error tolerances for projection errors are $10^{-11}$ for all the linear bases and $3.6 \times 10^{-16}$, $4.4 \times 10^{-16}$, $5.1 \times 10^{-16}$, and $1.0 \times 10^{-15}$ for quadratic bases with $T=512\,\mathrm{s}$, $256\,\mathrm{s}$, $128\,\mathrm{s}$, and $64\,\mathrm{s}$ respectively. For the \ac{bns} \phenompnrtidal{} bases, they are $10^{-12}$ for all the linear bases and $3.6 \times 10^{-17}$, $4.4 \times 10^{-17}$, $5.1 \times 10^{-17}$, and $1.0 \times 10^{-16}$ for quadratic bases with $T=512\,\mathrm{s}$, $256\,\mathrm{s}$, $128\,\mathrm{s}$, and $64\,\mathrm{s}$ respectively. The lower tolerance values for \phenompnrtidal{} is to avoid any systematic biases from the \ac{roq} approximation on the measurement of tidal effects. For the \ac{nsbh} \phenomp{} bases, the error tolerances for projection errors are $10^{-10}$ for all the linear bases and $1.4 \times 10^{-14}$, $1.7 \times 10^{-14}$, $2.0 \times 10^{-14}$, $6.2 \times 10^{-15}$, and $1.2 \times 10^{-14}$ for quadratic bases with $T=128\,\mathrm{s}$, $64\,\mathrm{s}$, $32\,\mathrm{s}$, $16\,\mathrm{s}$, and $8\,\mathrm{s}$ respectively.

\subsection{Basis sizes and speed-up gains} \label{sec:basis_sizes}

\begin{table*}
\centering
\begin{tabular}{c | c c | c c c | c c | c c c}

\multirow{2}{*}{Waveform} & 
\multicolumn{2}{c |}{$\mathcal{M}~(M_\odot)$} &
\multicolumn{3}{c |}{Frequencies (Hz)} & 
\multicolumn{2}{c |}{\#Bases} &
\multicolumn{2}{c}{Basis size} &
\multirow{2}{*}{Speedup} \\

 & Min & Max &
$f_{\mathrm{low}}$ & $f_{\mathrm{high}}$ & $1 / T$ &
Linear & Quadratic &
Linear & Quadratic & \\

\hline

\multirow{4}{*}{\phenomd{}} & $0.6$ & $1.1$ & $20$ & $1024$ & $1/512$ & $149$ & $1$ & $126$--$137$ & $24$ & $250$--$460$ \\
& $0.92$ & $1.7$ & $20$ & $1024$ & $1/256$ & $74$ & $1$ & $120$--$130$ & $25$ & $110$--$210$ \\
& $1.4$ & $2.6$ & $20$ & $1024$ & $1/128$ & $37$ & $1$ & $112$--$122$ & $28$ & $58$--$100$ \\
& $2.1$ & $4.0$ & $20$ & $1024$ & $1/64$ & $20$ & $1$ & $109$--$117$ & $32$ & $29$--$43$ \\

\hline

\multirow{4}{*}{\phenomp{} (BNS)} & $0.6$ & $1.1$ & $20$ & $4096$ & $1/512$ & $15$ & $15$ & $639$--$788$ & $454$--$646$ & $790$--$1100$ \\
& $0.92$ & $1.7$ & $20$ & $4096$ & $1/256$ & $8$ & $8$ & $567$--$633$ & $380$--$491$ & $460$--$550$ \\
& $1.4$ & $2.6$ & $20$ & $4096$ & $1/128$ & $4$ & $4$ & $555$--$567$ & $335$--$392$ & $260$--$300$ \\
& $2.1$ & $4.0$ & $20$ & $2048$ & $1/64$ & $2$ & $2$ & $526$--$527$ & $291$--$308$ & $69$--$70$ \\

\hline

\multirow{4}{*}{\phenompnrtidal{}} & $0.6$ & $1.1$ & $20$ & $4096$ & $1/512$ & $15$ & $15$ & $861$--$964$ & $543$--$741$ & $830$--$990$ \\
& $0.92$ & $1.7$ & $20$ & $4096$ & $1/256$ & $8$ & $8$ & $803$--$859$ & $487$--$587$ & $450$--$550$ \\
& $1.4$ & $2.6$ & $20$ & $4096$ & $1/128$ & $4$ & $4$ & $769$--$813$ & $466$--$508$ & $230$--$280$ \\
& $2.1$ & $4.0$ & $20$ & $2048$ & $1/64$ & $2$ & $2$ & $756$--$765$ & $457$--$466$ & $58$--$60$ \\

\hline

\multirow{5}{*}{\phenomp{} (Intermediate)} & $1.4$ & $2.6$ & $20$ & $1024$ & $1/128$ & $37$ & $37$ & $1540$--$1824$ & $2808$--$3484$ & $26$--$31$ \\
& $2.1$ & $4.0$ & $20$ & $1024$ & $1/64$ & $20$ & $20$ & $1716$--$2092$ & $3256$--$4034$ & $11$--$14$ \\
& $3.3$ & $6.3$ & $20$ & $1024$ & $1/32$ & $10$ & $10$ & $1979$--$2377$ & $3840$--$4170$ & $5.7$--$6.4$ \\
& $5.2$ & $11.0$ & $20$ & $1024$ & $1/16$ & $5$ & $5$ & $2304$--$2404$ & $3794$--$4056$ & $3.0$--$3.2$ \\
& $8.7$ & $21.0$ & $20$ & $1024$ & $1/8$ & $3$ & $3$ & $2062$--$2232$ & $2525$--$3204$ & $2.0$--$2.3$ \\

\hline

\multirow{4}{*}{\phenomxphm{}} & $10.02$ & $19.05$ & $20$ & $4096$ & $1/32$ & $23$ & $23$ & $1870$--$2092$ & $2444$--$2619$ & $14$--$19$ \\
& $15.52$ & $31.85$ & $20$ & $4096$ & $1/16$ & $15$ & $15$ & $1886$--$3095$ & $2463$--$3095$ & $4.9$--$8.0$ \\
& $26.54$ & $62.86$ & $20$ & $4096$ & $1/8$ & $11$ & $11$ & $1222$--$1836$ & $1222$--$1836$ & $2.4$--$3.8$ \\
& $52.38$ & $200.0$ & $20$ & $4096$ & $1/4$ & $11$ & $11$ & $308$--$806$ & $308$--$806$ & $1.9$--$3.7$

\end{tabular}
\caption{Chirp-mass partitions, frequency range, number of bases for each partition, basis sizes and speed-up gains of our \ac{roq} bases. See Sec. \ref{sec:mass_partitions} for how the chirp-mass partitions are determined and Sec. \ref{sec:subdomain} for how each partition is divided into chirp-mass sub-domains, for each of which \ac{roq} basis is constructed. The speed-up gains are measured speed-up gains in likelihood evaluations. See Sec. \ref{sec:basis_sizes} for how they are measured.}
\label{tab:basis_size}
\end{table*}

The sizes and speed-up gains of the bases are presented in Tab. \ref{tab:basis_size}. The low-spin \phenomd{} bases are most compact and have only $\sim 100$ basis elements for each in total. The \phenomp{} and \phenompnrtidal{} bases of the \ac{bns} mass range have several hundreds to $\sim 1000$ basis elements for each, and the latter ones are larger due to the presence of tidal effects. The \phenomp{} bases of the intermediate mass range have higher basis sizes (up to several thousands) than the \ac{bns} bases due to the extended range of mass ratio. The sizes of the \phenomxphm{} bases range from several hundreds to thousands and significantly depend on chirp mass range. 

The speed-up gains presented in the tables are measured speed-up gains in likelihood evaluations. For each chirp mass sub-domain, log-likelihood-ratio is evaluated for 1000 random source parameter samples with and without the \ac{roq} approximation and a speed-up gain is calculated as the ratio of evaluation time for each sample. Each row in the table presents the range of the medians of measured speed-up gains. For the low-spin \phenomd{} bases, the measured speed-up gains are $\sim 10$ times lower than the expected speed-up gains $K / (N_L + N_Q)$. It arises because waveform evaluations with those bases are so quick that other fixed cost such as pre-computations of \phenomd{} amplitude and phase coefficients dominate the cost. For the other cases, the measured speed-up gains are roughly same as the expected gains.

\subsection{Likelihood errors} \label{sec:likelihood_errors}

\begin{figure}[t]
	\centering
    	\includegraphics[width=0.98\linewidth]{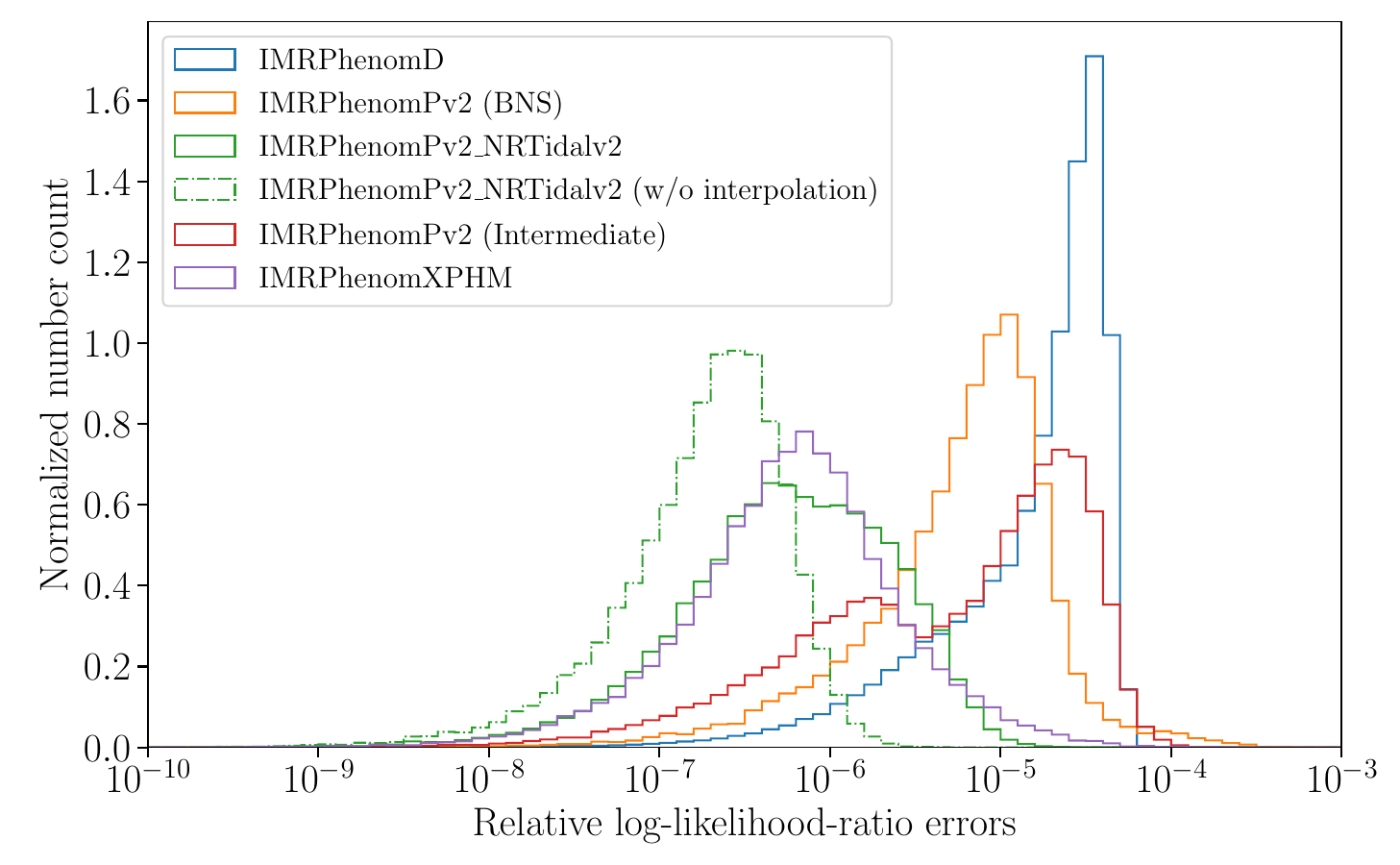}
    	\caption{Relative log-likelihood-ratio errors of our ROQ bases for random source parameter samples. They have been computed for hundreds of source parameter samples per basis sub-domain.} \label{fig:likelihooderrors}
\end{figure}

Figure \ref{fig:likelihooderrors} presents relative log-likelihood-ratio errors introduced by our \ac{roq} bases. The errors have been computed for hundreds of source parameter samples per sub-domain. For each parameter sample $\theta$, a simulated signal is generated and considered as observed data $d_i(f) = h_i(f;\theta)$, and the relative error between $\ln \Lambda (\{d_i\}_{i=1}^{N_{\mathrm{det}}}|\theta)$ and $\ln \Lambda_{\mathrm{ROQ}} (\{d_i\}_{i=1}^{N_{\mathrm{det}}}|\theta)$ is computed. We consider only a single LIGO detector with its design sensitivity for this study. The source parameters contain amplitude and phase calibration errors at 10 log-uniformly distributed frequency nodes, and interpolated calibration errors are multiplied by simulated signal and template waveform used for likelihood evaluations. Their values at the nodes are drawn from uniform distribution from $-20\%$ to $20\%$ for amplitude and $-15\,\mathrm{deg}$ to $15\,\mathrm{deg}$ for phase.  

As seen in the figure, the relative errors are $\lesssim 10^{-5}$ for the \phenompnrtidal{} bases and $\lesssim 10^{-4}$ for the other bases. The lower errors for \phenompnrtidal{} are due to the tighter error tolerances explained in the previous sub-section. Since log-likelihood-ratio is in the order of squared \ac{snr}, the absolute log-likelihood-ratio errors are smaller than unity for $\text{\ac{snr}} < 100$, and our \ac{roq} bases will not introduce biases in inference for typical \ac{snr} values observed by \ac{lvk} detectors. We also note that the dominant errors of \phenompnrtidal{} do not come from the bases themselves, but the numerical interpolation of \ac{roq} weights over the coalescence time $t_{\mathrm{c}}$. Relative errors computed when weights are calculated exactly are shown as the dashed-dotted line in the figure, and the errors get reduced to $\lesssim 10^{-6}$. 

\section{Applications} \label{sec:application}

In this section, we demonstrate the usefulness of our rapid parameter estimation framework in various applications. 

\subsection{Rapid localization of \ac{bns}}

Our rapid parameter estimation framework can be applied to rapid and accurate source localization of \ac{bns} signals for use in searches for their electromagnetic counterparts.
In the current \ac{lvk} alert system, the rapid source localization software, \bayestar{} \cite{Singer:2015ema, Singer:2016eax}, is run once \ac{cbc} signal is detected.
It utilizes output from a \ac{cbc} search pipeline performing the matched-filtering~\cite{Wainstein1970, Sathyaprakash:1991aa, Finn:1992aa, Finn:1993aa} process on strain data.
Specifically, it reads in matched-filter \ac{snr} time series for each detector, and calculates the posterior probability distribution over sky location and luminosity distance to the source with a run time of seconds.
The input matched-filter \ac{snr} time series is computed with the best-matching template included in \textit{a template bank}, a collection of simulated gravitational waveforms for various mass and spin values over which matched-filtering process is performed.
To mitigate potential bias or loss of precision due to the mismatch between the signal and the best-matching template, \bilby{} is run to explore mass--spin space broader than that covered by the template bank and update the estimate of source location.
Our low-spin \phenomd{} \ac{roq} bases have been developed specifically for speeding up this update procedure for \ac{bns} signal.

In this section we benchmark the speed and localization accuracy of \bilby{} parameter estimation using the low-spin \phenomd{} \ac{roq} bases.
We inject simulated signals, commonly called \textit{injections}, into \ac{o3} data of the \ac{hlv} detector network, which are publicly available \cite{LIGOScientific:2023vdi}.
Then, we recover their locations with our rapid parameter estimation framework, and investigate the run time and recovery accuracy.
To benchmark its performance in a realistic situation, we perform end-to-end tests, where the simulated data are analyzed by a search pipeline and parameter estimation is run with the settings determined based on the search outputs.
We also run \bayestar{} on the injections for comparison.

The source parameters of the injections are randomly drawn from the astrophysical distribution we assume.
The detector-frame component masses are drawn from a uniform distribution across $1 M_\odot \leq m_1, m_2 \leq 3 M_\odot$.
The spins are assumed to be parallel with the orbital angular momentum, and their components projected over the orbital angular momentum are drawn from a uniform distribution in $-0.05 \leq \chi_1, \chi_2 \leq 0.05$.
The effects of tidal deformation of colliding bodies are not taken into account in injections.
The source locations are distributed isotropically over sky location and uniformly in the cubic of luminosity distance $D_\mathrm{L}$ between $30\,\mathrm{Mpc} \leq D_\mathrm{L}  \leq 600\,\mathrm{Mpc}$.
The distribution is isotropic in binary orientation and uniform over coalescence phase of binary motion.
To avoid a lot of injections whose optimal \ac{snr}{s} are too small to detect, we pre-estimate the network optimal \ac{snr}, the root-mean-square of optimal \ac{snr}{s} at all the detectors, of each signal using reference \ac{o3} detector sensitivities, and inject only signals whose network optimal \ac{snr}{s} exceed $8 \sqrt{2}$\footnote{\ac{snr} of 8 is a typical threshold used to estimate the observable range of a detector, and \ac{snr}{s} exceeding 8 coincidentally at two detectors requires the network \ac{snr} to be larger than $8 \sqrt{2}$.}, yielding 1047 injections in total.
Those signals are synthesized based on the \phenomd{} waveform model and injected into the \ac{o3} dataset between 13 June 2019 18:46 UTC and 16 August 2019 12:45 UTC.
They are placed so that the interval of coalescence time $\tc$ between neighboring injections is longer than $100\,\mathrm{s}$ to mitigate biases due to signal overlap.\footnote{We have found that only one pair of neighboring injections have time-frequency overlap, where the latter injection has lower masses than those of the earlier one and has longer signal duration. We expect its effects on our main results are negligible, since they are statistical results from hundreds of injections and the overlap has limited effects on the frequency integral of likelihood.} 

To simulate a gravitational-wave search, we use the \gstlal{} search pipeline (referred to as \gstlal{} hereafter) \cite{Cannon:2020qnf} with a template bank constructed based on a stochastic placement algorithm~\cite{template_placement1, template_placement2}.
Each template is generated using the \tf{} waveform model \cite{Buonanno:2009zt}.
For quick tests, we apply the matched-filter \ac{snr} maximized over coalescence phase and time as detection statistics rather than performing full likelihood analysis. 
Specifically, we apply the network matched-filter \ac{snr} above 12 and second largest \ac{snr} among the $3$ detectors above 5.5 as the detection threshold, recovering 481 injections in total.

For 308 injections, data from all the $3$ detectors are available since all of the detectors were in observing mode.
For the other 173 injections, one of the detectors were not in the observing mode.
We refer to those two types of injections as \textit{triple-detector injections} and \textit{double-detector injections} respectively.
For each injection, the detection criteria are typically met for multiple templates in the template bank.
We refer to the template with the highest \ac{snr} among them as \textit{preferred template}, and the matched SNR time series computed with that template is used as input to \bayestar{}.

For \bilby{} analysis, we employ the \dynesty{} sampler \cite{Speagle:2019ivv} with the acceptance-walk \ac{mcmc} method.
The number of live points and the average number of accepted \ac{mcmc} jumps are set to $500$ and $10$ respectively.
The sampling is parallelized with 24 processes.
For each simulated signal, two independent runs are performed and their samples are combined.
The \ac{psd} produced by \gstlal{} is used for likelihood evaluations.

The prior probability distribution is the same as that used to populate the injections, except for the explored range of $\mathcal{M}$.
Since the chirp mass of the preferred template $\mathcal{M_{\mathrm{template}}}$ is typically very close to its true value for \ac{bns} \cite{Biscoveanu:2019ugx}, its explored range is set to $0.995 \mathcal{M_{\mathrm{template}}} \leq \mathcal{M} \leq 1.005\mathcal{M_{\mathrm{template}}}$.
The explored range determined in that way includes the true chirp mass value except for 3 double-detector and 4 triple-detector injections. 
The errors of chirp mass recoveries for those 7 injections are higher than $20\%$ and much higher than the errors for the other detections. 
Hence we anticipate it is due to non-stationary or non-Gaussian noise around the injections.
We exclude those 7 injections in the main results presented in this section, while the effects of their inclusion are briefly explained in the text.
We analytically marginalize over the coalescence phase and we marginalize over the luminosity distance using the look-up table method, and they are recovered at the post processing stage.
Detector calibration uncertainties are not taken into account in these simulations and are not marginalized over.

\begin{figure}[t]
	\centering
    	\includegraphics[width=0.98\linewidth]{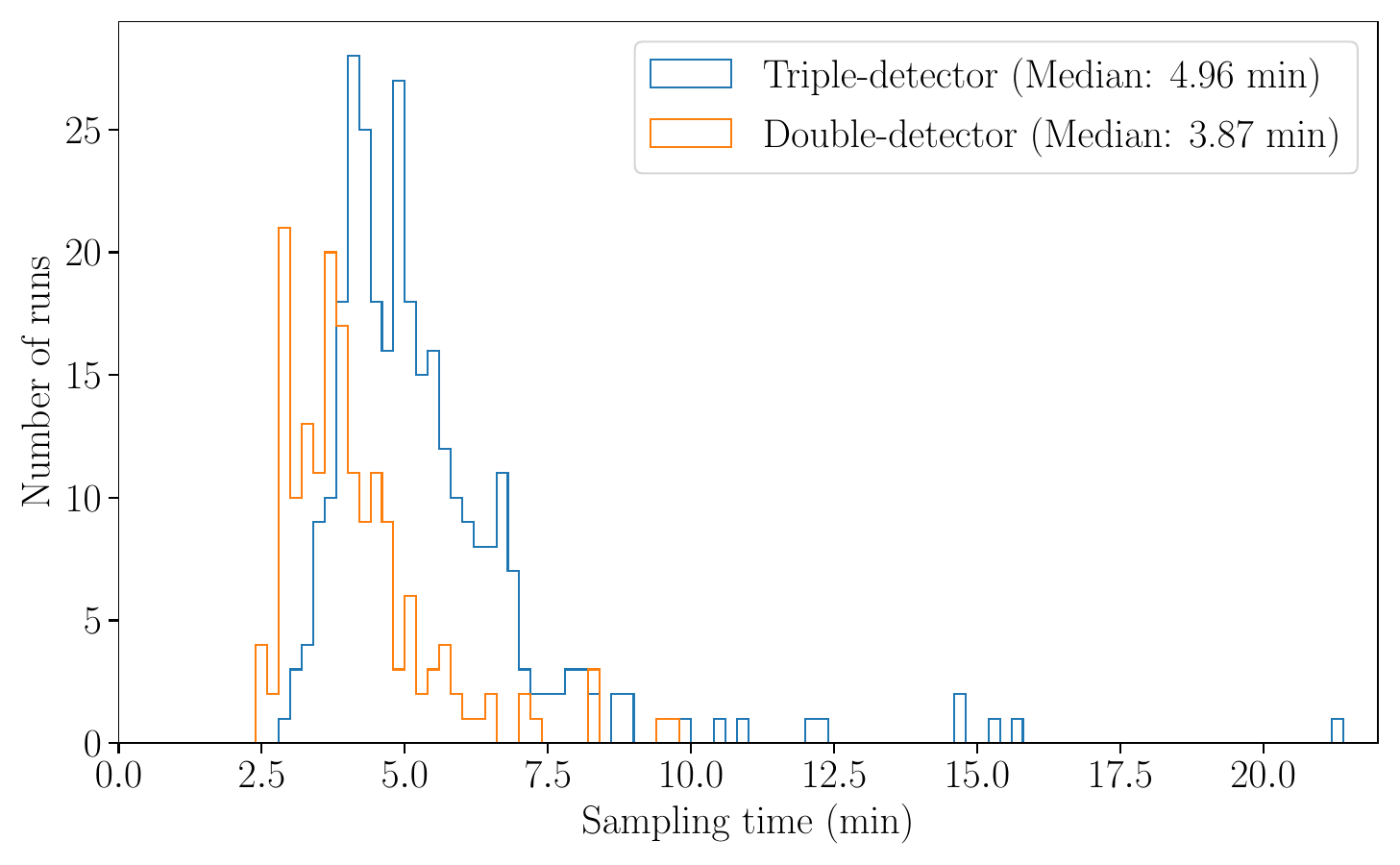}
    	\caption{Sampling time of \bilby{} with the low-spin \phenomd{} \ac{roq} bases for simulated \ac{bns} signals. The blue and orange histograms show sampling time for 304 triple-detector and 170 double-detector injections respectively, and the median sampling time is presented in the legend.} \label{fig:samplingtime}
\end{figure}

\begin{figure*}[t]
	\centering
	\begin{minipage}{0.49\linewidth} 
		\centering
		\subfloat[triple-detector injections]{
			\includegraphics[width=0.9\linewidth]{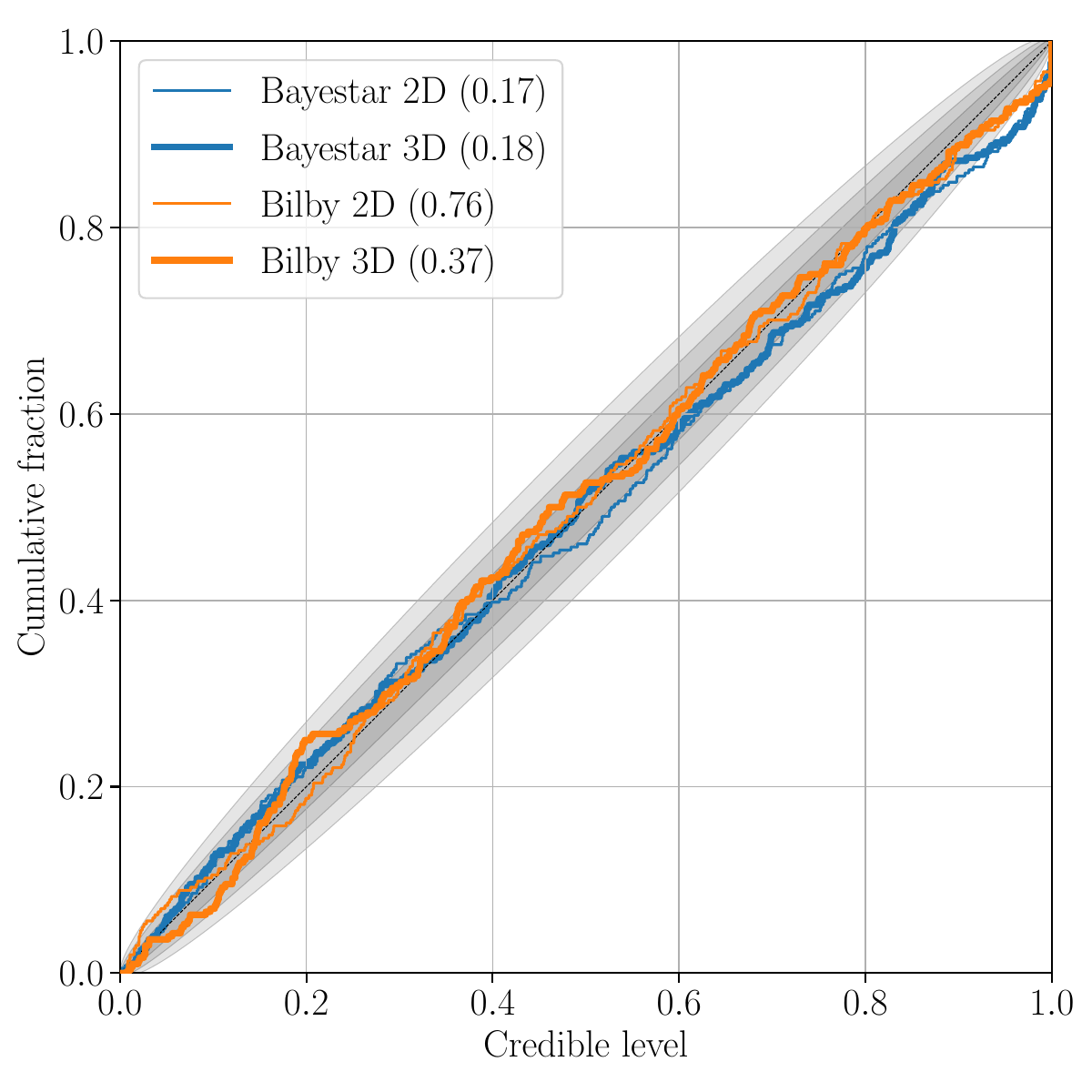}
		}
	\end{minipage}
	\begin{minipage}{0.49\linewidth} 
		\centering
		\subfloat[double-detector injections]{
			\includegraphics[width=0.9\linewidth]{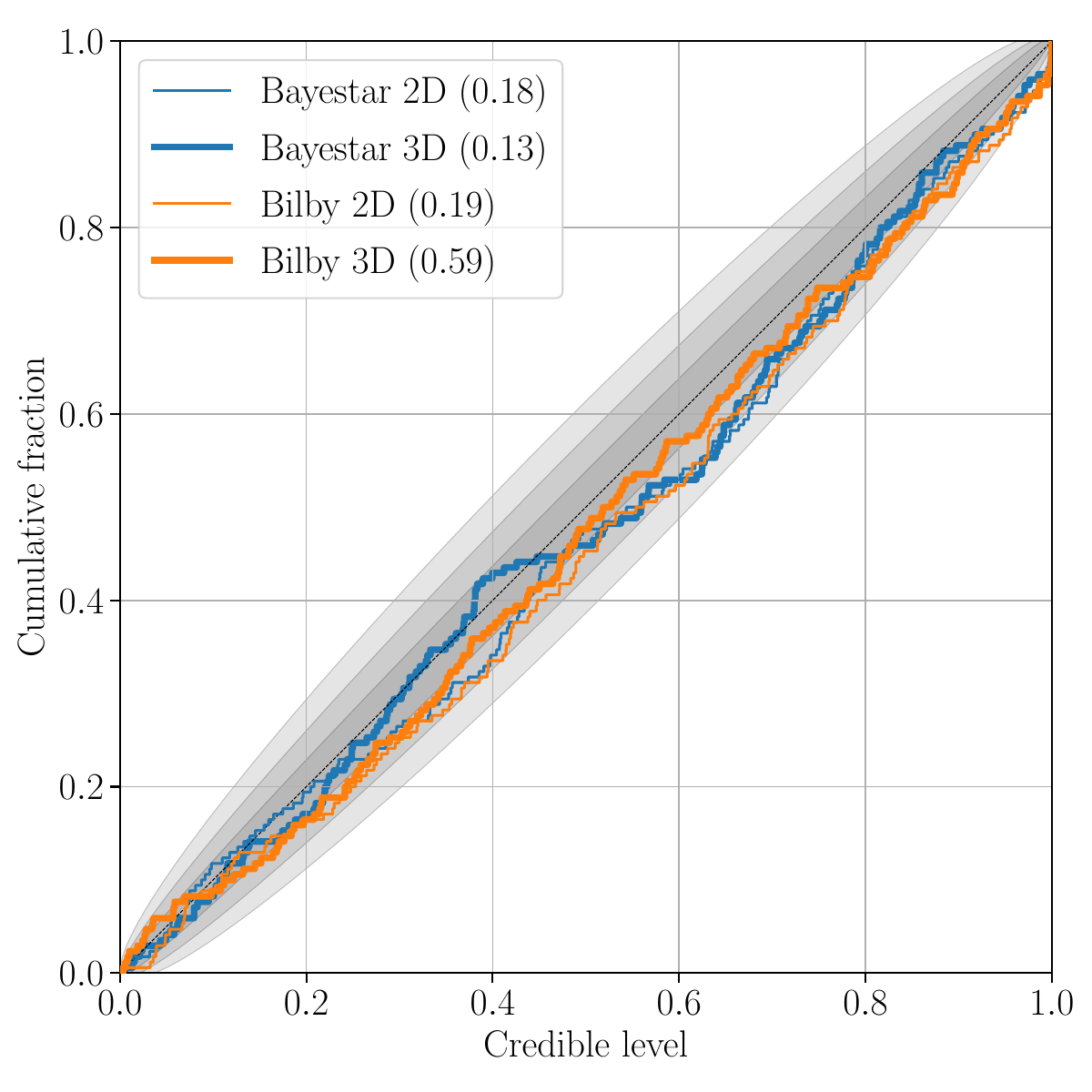}
		}
	\end{minipage} 
	\caption{$P$-$P$ plots of estimates on sky location (2D, thin line) and 3-dimensional location including distance (3D, thick line) from \bayestar{} (blue) and \bilby{} (orange) running on simulated \ac{bns} signals. The left and right panels show results from 304 triple-detector and 170 double-detector injections respectively. The gray bands represent the $1$, $2$, and $3$-$\sigma$ quantiles of statistical errors due to the finite number of samples. The $p$-value of \ac{ks} test between the observed credible levels and a uniform distribution for each case is presented in the legend.} \label{fig:pp}
\end{figure*}

\begin{figure*}[t]
	\centering
	\begin{minipage}{0.49\linewidth} 
		\centering
		\subfloat[searched areas for triple-detector injections]{
			\includegraphics[width=0.95\linewidth]{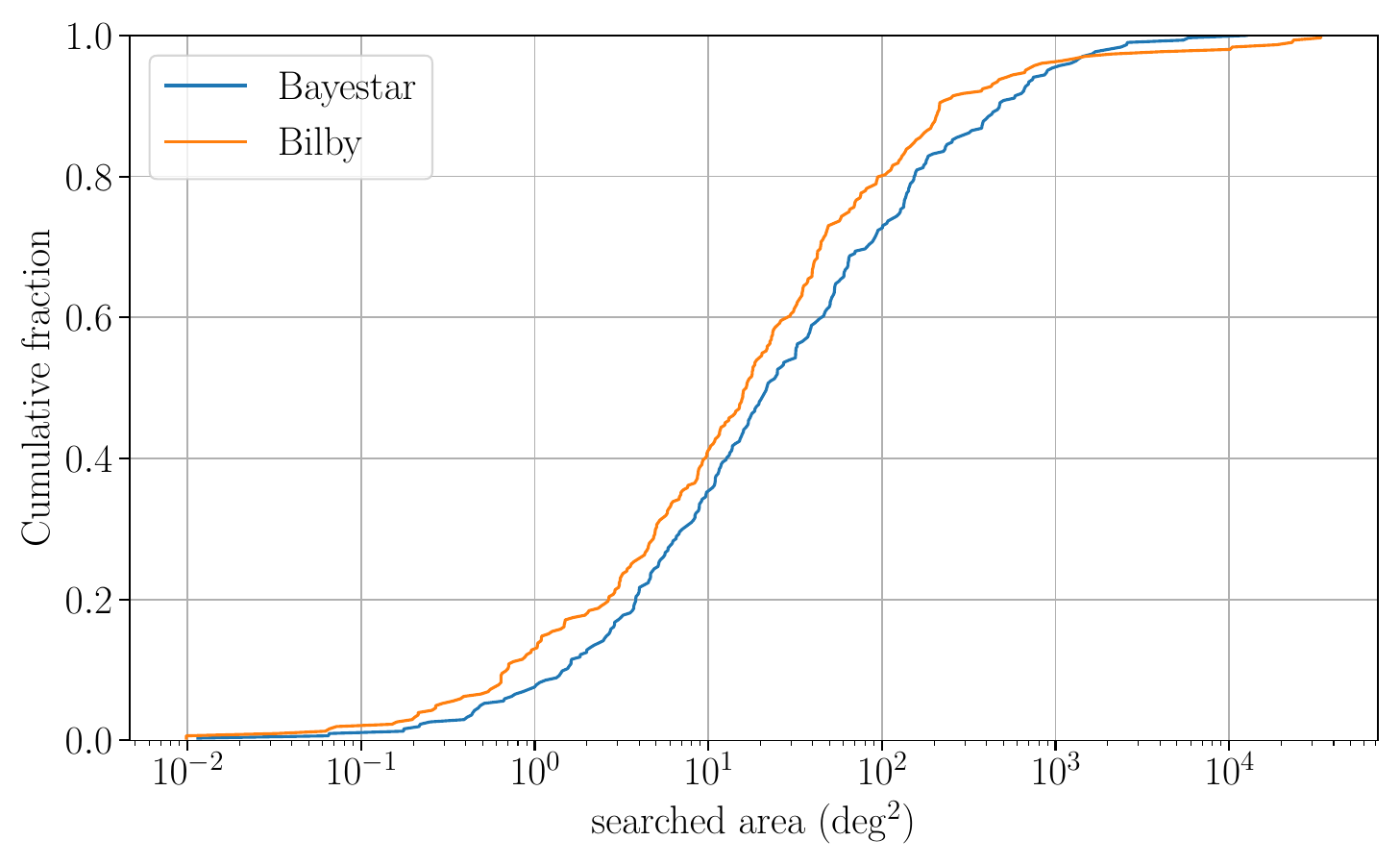}
		}
	\end{minipage}
	\begin{minipage}{0.49\linewidth} 
		\centering
		\subfloat[searched volumes for triple-detector injections]{
			\includegraphics[width=0.95\linewidth]{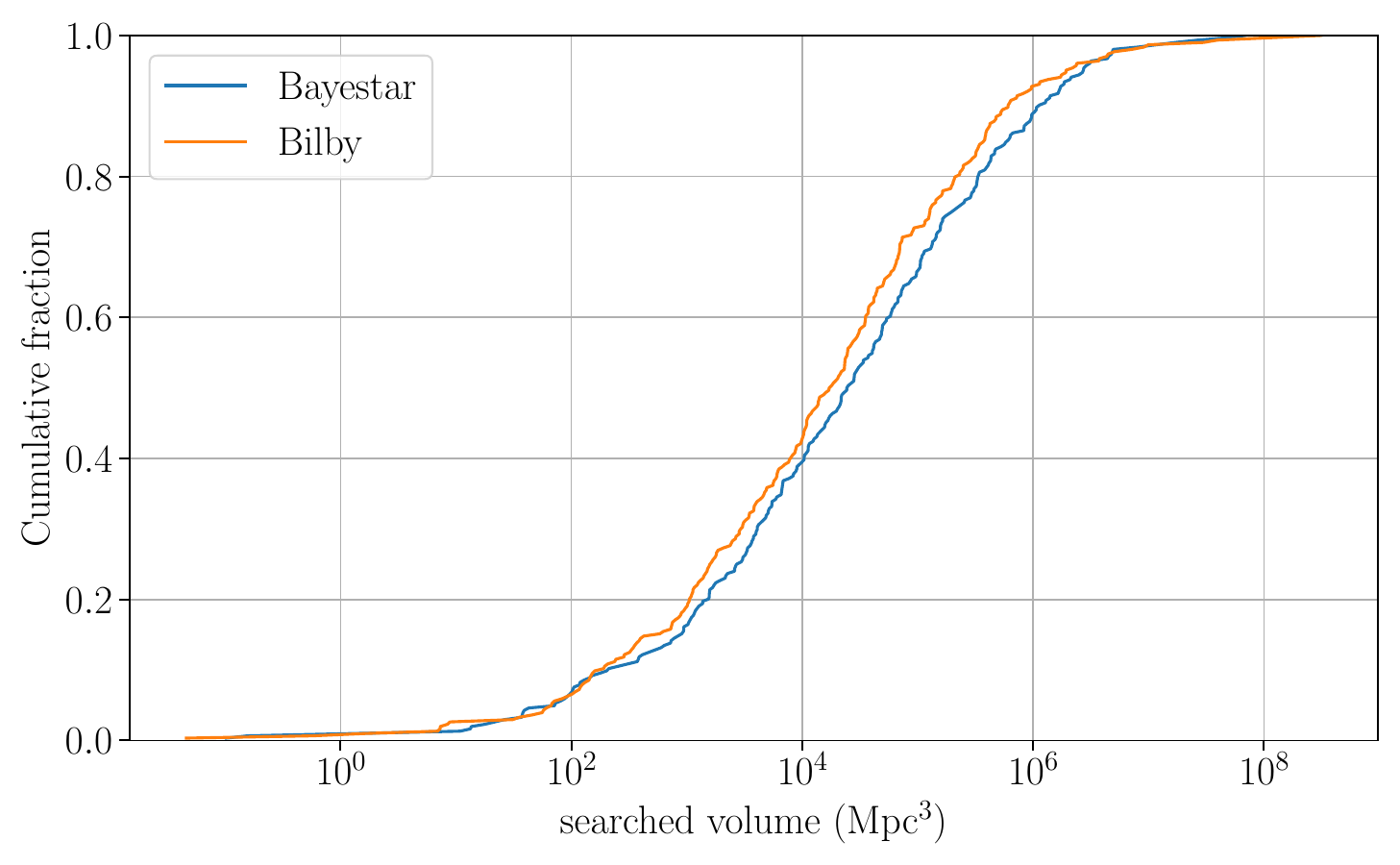}
		}
	\end{minipage} \\
	\begin{minipage}{0.49\linewidth} 
		\centering
		\subfloat[searched areas for double-detector injections]{
			\includegraphics[width=0.95\linewidth]{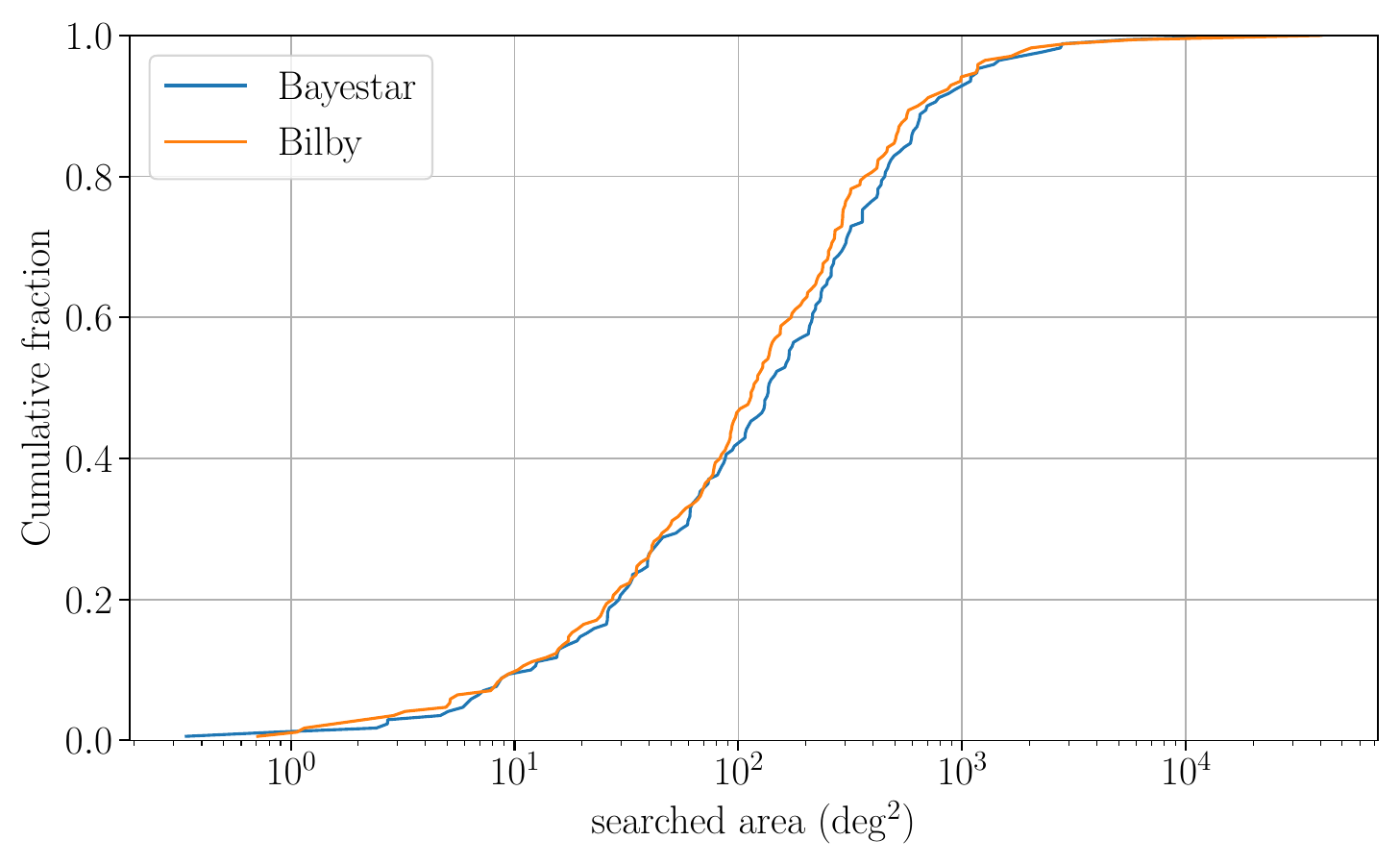}
		}
	\end{minipage}
	\begin{minipage}{0.49\linewidth} 
		\centering
		\subfloat[searched volumes for double-detector injections]{
			\includegraphics[width=0.95\linewidth]{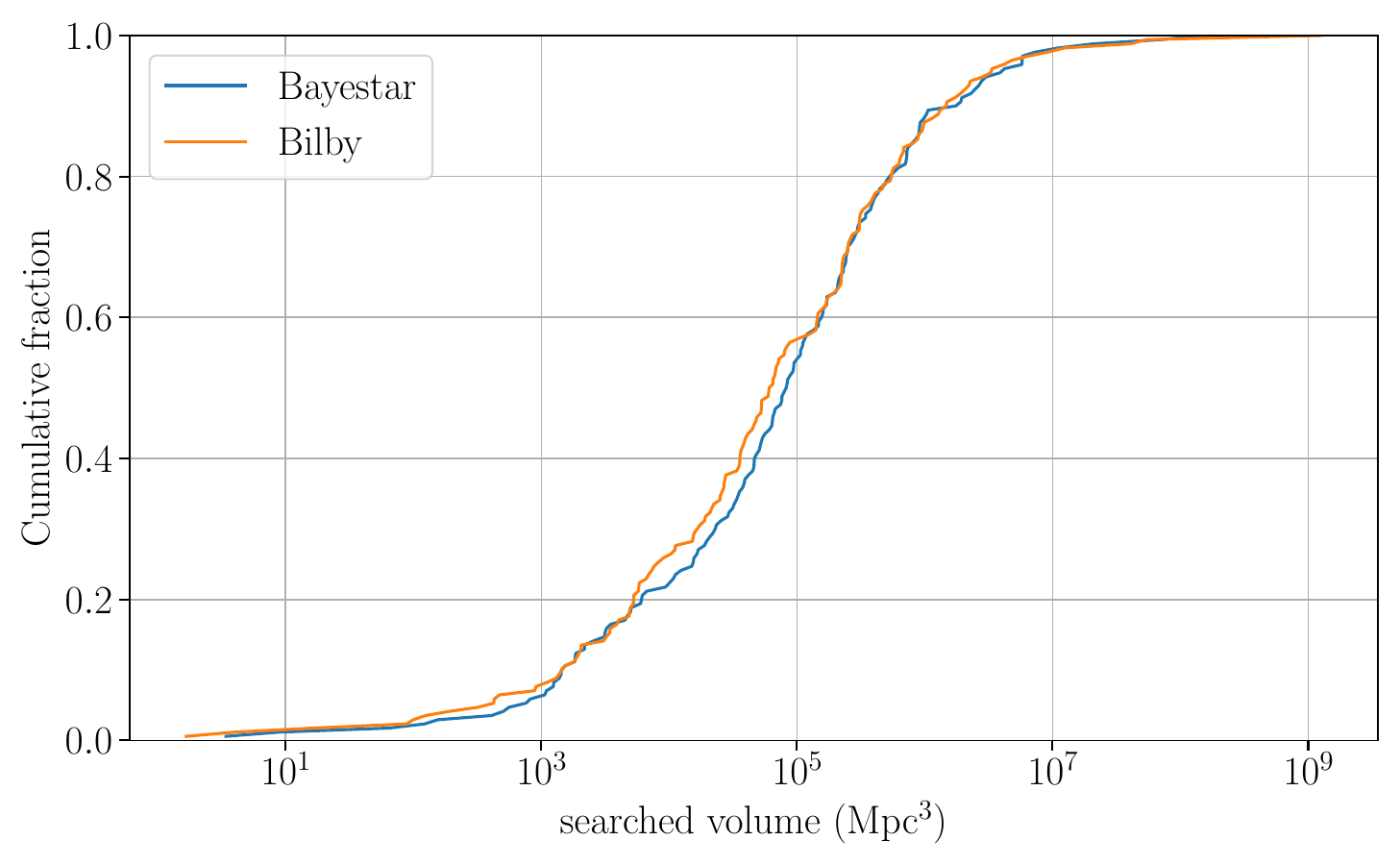}
		}
	\end{minipage}
	\caption{Cumulative distribution of searched areas and volumes from \bayestar{} (blue) and \bilby{} (orange) running on simulated \ac{bns} signals. The top and bottom panels show searched areas and volumes respectively from 304 triple-detector (left) and 170 double-detector (right) injections. By updating \bayestar{} results with \bilby{} results, the median searched area (volume) is reduced from $21.8\,\mathrm{deg^2}$ ($2.47 \times 10^4\,\mathrm{Mpc^3}$) to $16.6\,\mathrm{deg^2}$ ($1.75 \times 10^4\,\mathrm{Mpc^3}$) for the triple-detector case and $137\,\mathrm{deg^2}$ ($8.30 \times 10^4\,\mathrm{Mpc^3}$) to $117\,\mathrm{deg^2}$ ($6.27 \times 10^4\,\mathrm{Mpc^3}$) for the double-detector case.} \label{fig:searched}
\end{figure*}

Figure \ref{fig:samplingtime} presents the histogram of time taken by \bilby{} sampling for each signal.
As seen in the figure, most of the runs complete within several minutes.
The median run time is $4.96\,\mathrm{min}$ for the triple-detector case and $3.87\,\mathrm{min}$ for the double-detector case.
The runs are faster for the double-detector case since less operations are required to evaluate the likelihood due to fewer detectors.
The runs are performed with an Intel Xeon Gold 6136 CPU with a clock rate of 3.0 GHz.
In addition to the sampling time, pre-computations of \ac{roq} weights take $\sim 1\,\mathrm{min}$ for the triple-detector case and less for the double-detector case.
Thus, the total run time is minutes to 10 minutes, which provides enough time for follow-up observations of optical radiation fading away with the time scale of days \cite{LIGOScientific:2017ync}.

Figure \ref{fig:pp} presents $P$-$P$ plots of sky location (2D) and 3-dimensional location including distance (3D) for \bayestar{} and \bilby{}.
They are visual tools to check whether the true signal parameters are found within the $X\%$ credible region $X\%$ of the time, i.e., it tests whether the posteriors have the correct statistical properties.
If this is the case, the cumulative distribution of observed credible levels should be a diagonal line with statistical errors due to a finite number of samples.
The gray bands represent the $1$, $2$, and $3$-$\sigma$ quantiles of statistical errors.
For the triple-detector case, both \bayestar{} and \bilby{} $P$-$P$ plots are within the error band for credible level of $\lesssim 0.9$, while for the larger credible level \bayestar{} $P$-$P$ plots go outside the band and \bilby{} performs better. 
The $p$-value of \ac{ks} test between the observed credible levels and a uniform distribution for each case is presented in the legend, and in all the cases, the $p$-values of \bilby{} are larger than those from \bayestar{}, implying \bilby{} produces more accurate results mitigating search biases.
If the 7 injections whose chirp mass recoveries at the detection stage are bad are included, the $p$-values for \bayestar{} 2D, \bayestar{} 3D, \bilby{} 2D, and \bilby{} 3D are degraded to $(0.06, 0.07, 0.41, 0.42)$ for the triple-detector case and $(0.09, 0.06, 0.10, 0.31)$ for the double-detector case.

Figure \ref{fig:searched} presents cumulative distribution of searched areas and volumes, the areas and volumes one needs to search over from the most probable location to the least until reaching the true source location.
As seen in the figure, searched areas and volumes are systematically smaller for \bilby{}, demonstrating \bilby{} can reduce the area and volume follow-up observers need to search over.
The median searched area (volume) is reduced from $21.8\,\mathrm{deg^2}$ ($2.47 \times 10^4\,\mathrm{Mpc^3}$) to $16.6\,\mathrm{deg^2}$ ($1.75 \times 10^4\,\mathrm{Mpc^3}$) for the triple-detector case and $137\,\mathrm{deg^2}$ ($8.30 \times 10^4\,\mathrm{Mpc^3}$) to $117\,\mathrm{deg^2}$ ($6.27 \times 10^4\,\mathrm{Mpc^3}$) for the double-detector case.
In summary, \bilby{} with our rapid parameter estimation technique improves the estimate on source location with the time scale of minutes.

\subsection{Full \ac{bns} parameter estimation}

\begin{figure}[t]
	\centering
    \includegraphics[width=0.98\linewidth]{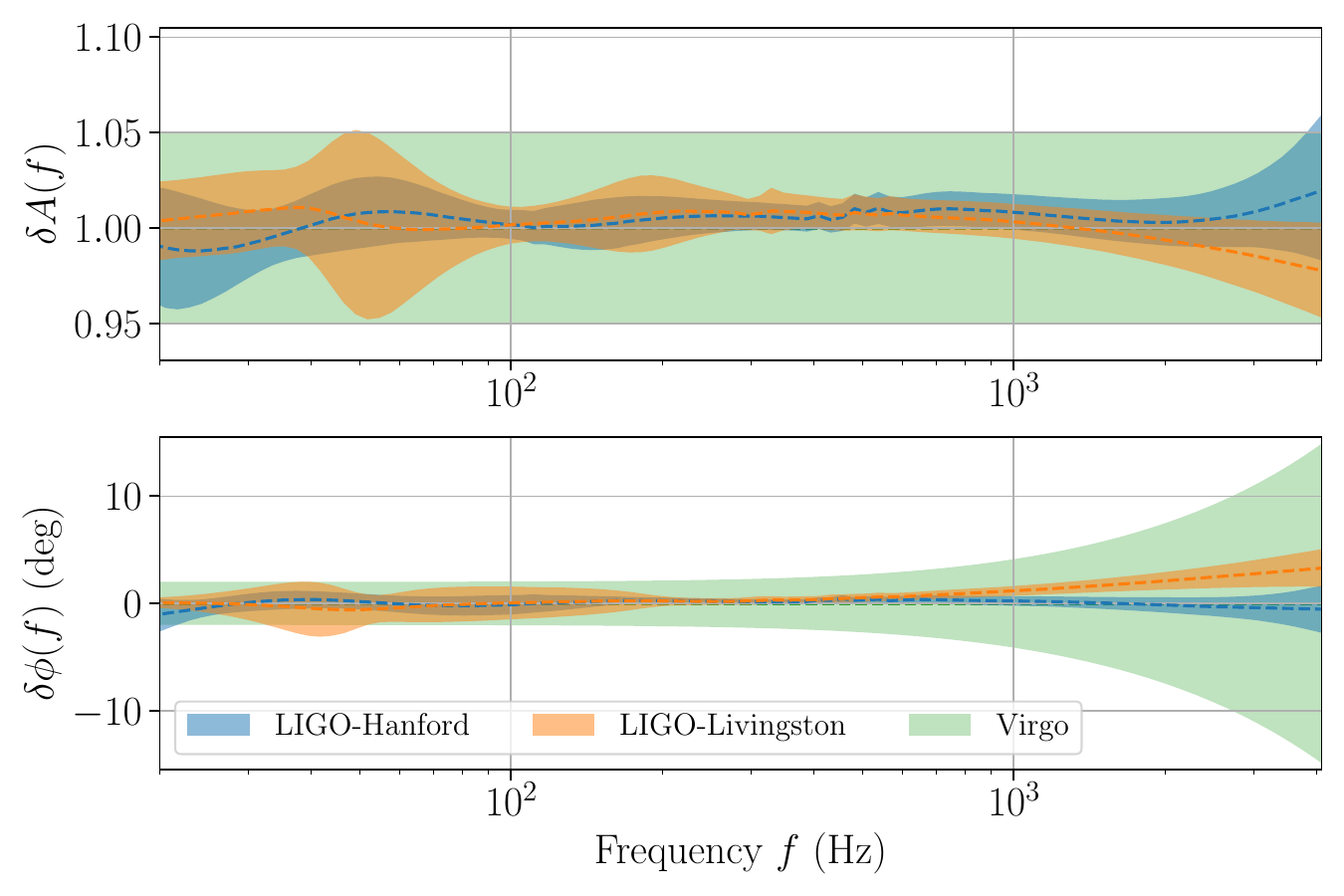}
	\caption{Detector calibration uncertainties of amplitude (top) and phase (bottom) used for simulations. The dashed lines represent the median values and the shaded regions represent the $1$-$\sigma$ uncertainties.} \label{fig:calibration}
\end{figure}

\begin{figure}[t]
	\centering
    \includegraphics[width=0.9\linewidth]{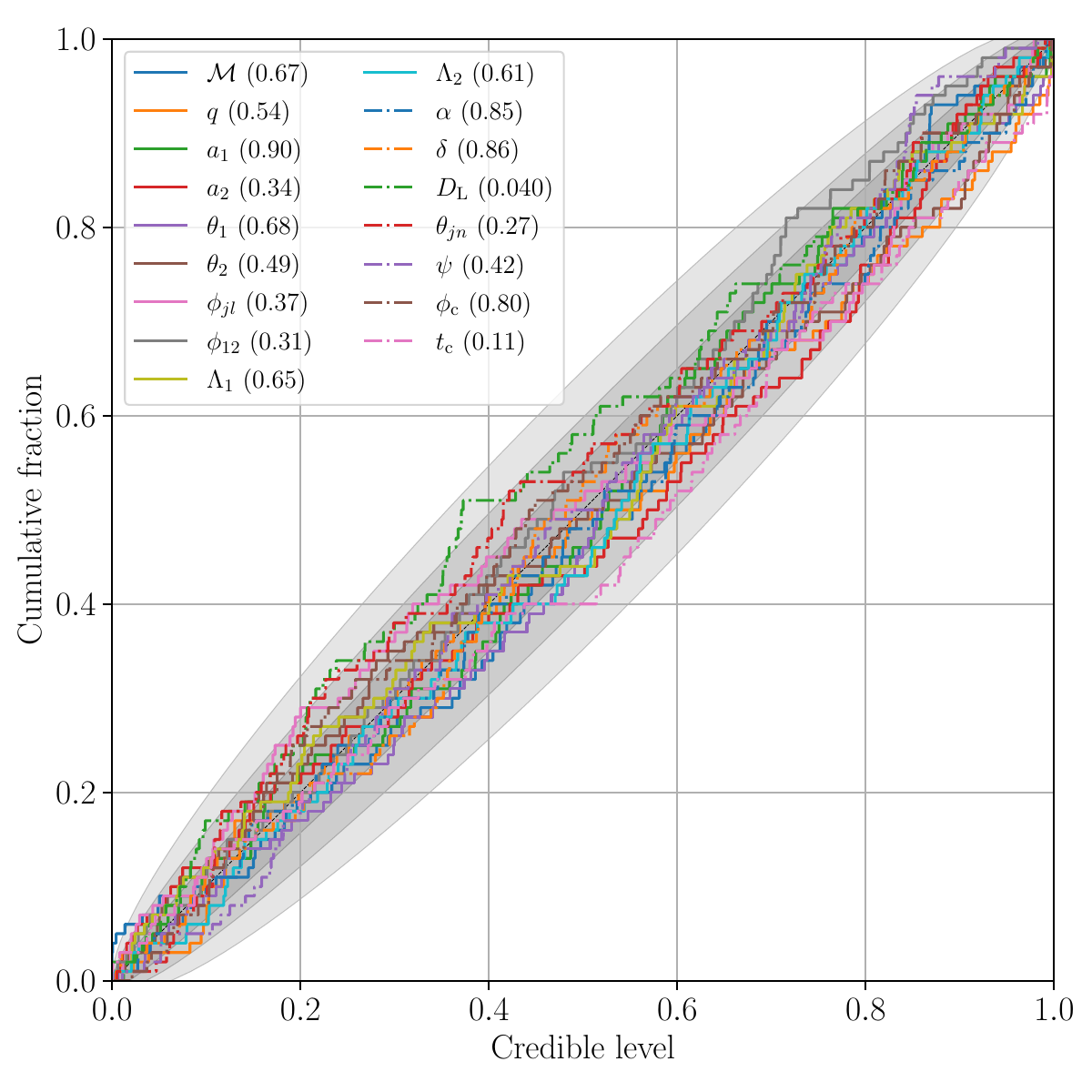}
	\caption{$P$-$P$ plots of all the parameters recovered by \bilby{} with the \phenompnrtidal{} \ac{roq} bases running on 100 injections. The recovered parameters are detector-frame chirp mass ($\mathcal{M}$), mass ratio ($q$), spin magnitudes ($a_1$ and $a_2$), spin angles ($\theta_1$, $\theta_2$, $\phi_{jl}$, and $\phi_{12}$), dimensionless tidal deformability parameters ($\Lambda_1$ and $\Lambda_2$), right ascension ($\alpha$) and declination ($\delta$) of the source location, luminosity distance to the source ($D_{\mathrm{L}}$), inclination angle between total angular momentum and line of sight ($\theta_{jn}$), polarization angle ($\psi$), coalescence phase ($\phi_{\mathrm{c}}$), and coalescence time ($t_{\mathrm{c}}$). The gray bands represent the $1$, $2$, and $3$-$\sigma$ quantiles of statistical errors due to the finite number of samples. The $p$-value of \ac{ks} test between the observed credible levels and a uniform distribution from $0$ to $1$ for each parameter is presented in the legend. The combined $p$-value, the probability that each $p$-value is drawn from a uniform distribution from $0$ to $1$, is 0.78.} \label{fig:pp_bns}
\end{figure}

We also demonstrate that our rapid parameter estimation framework can be applied to full \ac{bns} parameter estimation incorporating precession effects, tidal effects, and detector calibration uncertainties.
We inject 100 simulated \ac{bns} signals into simulated Gaussian noise data of a \ac{hlv} network drawn with their design sensitivities.
We then recover their source parameters using \bilby{} with our rapid parameter estimation framework and investigate its accuracy using the $P$-$P$ plots.

The injections are synthesized based on the \phenompnrtidal{} waveform model and analyzed with our \phenompnrtidal{} \ac{roq} bases of the $256$-$\mathrm{s}$ partition.
The detector-frame component masses are uniformly distributed with the chirp mass constraint $1.15M_\odot  \leq \mathcal{M}  \leq 1.25M_\odot  $ and the mass ratio constraint $q \geq 0.125$.
The spin magnitudes are uniformly distributed across $0 \leq a_1, a_2 \leq 0.4$ and the spin directions are isotropically distributed.
The dimensionless tidal deformability values are uniformly distributed in $0 \leq \Lambda_1,\Lambda_2 \leq 5000$.
The source locations are distributed uniformly in comoving volume and source frame time in the range $1\,\mathrm{Mpc} \leq D_{\mathrm{L}}  \leq 100\,\mathrm{Mpc}$.
The distribution is isotropic in binary orientation, uniform over coalescence phase, and uniform over coalescence time within a time window with the width of $0.2\,\mathrm{s}$.
The median network optimal \ac{snr} of injections is $21$.

Injections are multiplied by randomly generated calibration errors.
The amplitude and phase errors are calculated as cubic spline interpolation of their values at $10$ frequency nodes log-uniformly distributed from $20\,\mathrm{Hz}$ to $4096\,\mathrm{Hz}$.
We use calibration uncertainty budget of amplitude and phase at the GPS time of 1244415456 (12 June 2019 22:57:18 UTC), which is publicly available \cite{calibration} and presented in Fig. \ref{fig:calibration}.
The amplitude and phase errors at the nodes are drawn from Gaussian distribution whose mean matches the median shown as the dashed line and standard deviation is the half width of the $1$-$\sigma$ band shown as shaded region.

\bilby{} is run with the same prior probability distribution of source and calibration parameters as that to populate injections and the same \ac{psd} as that used for simulating Gaussian noise.
The \dynesty{} sampler is employed with the acceptance-walk \ac{mcmc} method, $500$ live points, and an average of $60$ accepted \ac{mcmc} jumps.
We analytically marginalize over the coalescence phase and we marginalize over the luminosity distance using the look-up table method, and they are recovered at the post processing stage.
The sampling is parallelized with 24 processes, and the median sampling time is $108\,\mathrm{min}$.
Without the \ac{roq} approximation, the expected sampling time is in the order of a month. 

Figure \ref{fig:pp_bns} presents $P$-$P$ plots of all the source parameters. 
As seen in the figure, they are well within the error band, and $p$-values in the legend imply the observed deviations are consistent with statical errors.
The combined $p$-value, the probability that each $p$-value is drawn from a uniform distribution from $0$ to $1$, is 0.78.
Hence, we conclude our rapid parameter estimation framework can also be applied to full and accurate parameter estimation analysis of \ac{bns} signal.

\subsection{BBH rapid parameter estimation}

\begin{figure}[t]
	\centering
    \includegraphics[width=0.9\linewidth]{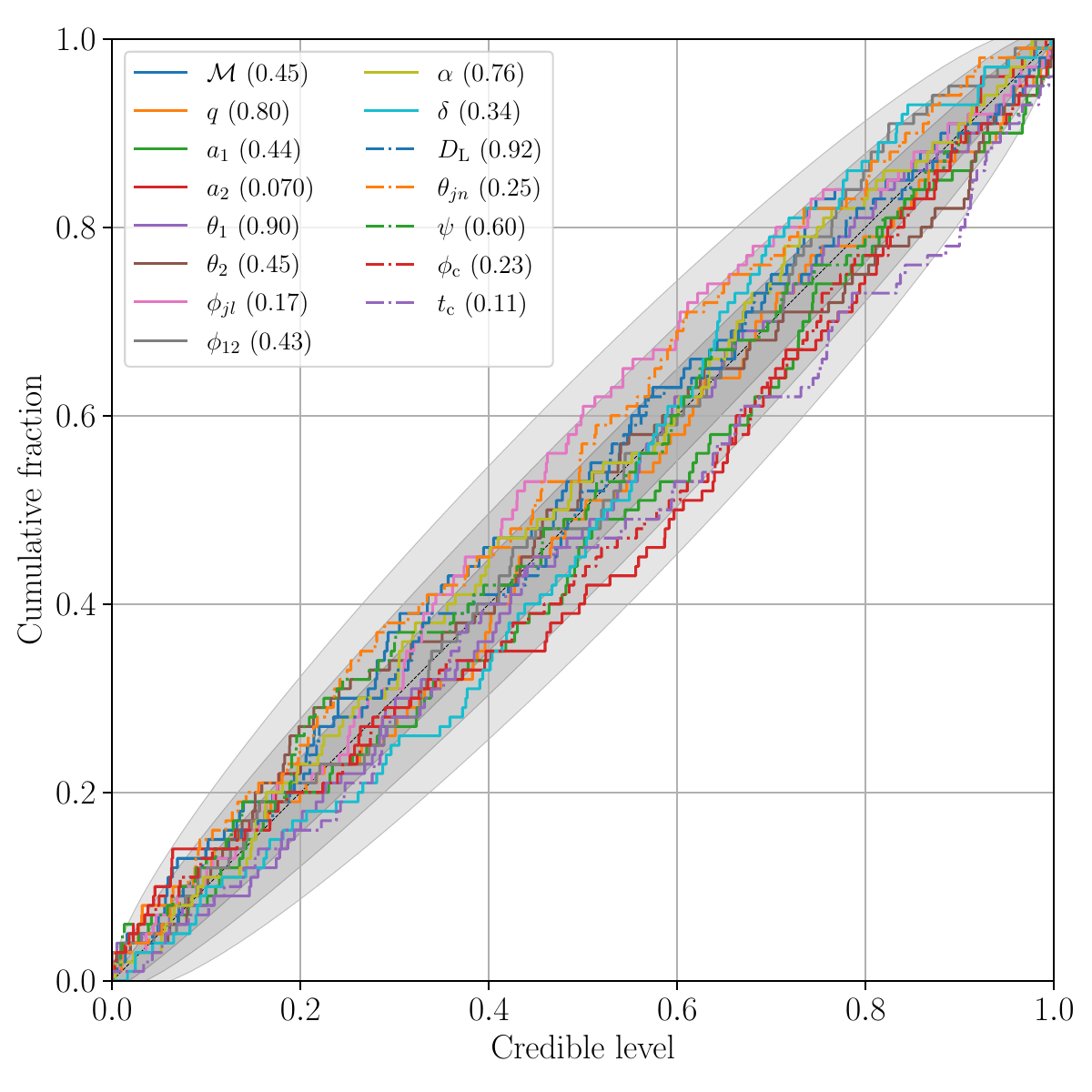}
	\caption{$P$-$P$ plots of all the parameters recovered by \bilby{} with the \phenomxphm{} \ac{roq} bases running on 100 injections. The combined $p$-value is 0.42.  See the caption of Fig. \ref{fig:pp_bns} for the definition of parameter symbols and the combined $p$-value, and the meaning of numbers in the legend and gray bands in the plot.} \label{fig:pp_bbh}
\end{figure}

Finally, we also demonstrate the application of the \phenomxphm{} bases to \ac{bbh} signals.
We inject 100 simulated \ac{bbh} signals synthesized based on the IMRPhenomXPHM waveform model into simulated Gaussian noise of a \ac{hlv} network and analyze them with our \phenomxphm{} $8$-s \ac{roq} bases.

The distribution to populate injections and analysis settings are almost same as what are used in the previous section with the following differences. 
The mass and spin range considered here is $26.54 M_\odot \leq \mathcal{M} \leq 62.86 M_\odot$, $1/20 \leq q \leq 1$, and $0 \leq a_1,a_2 \leq 0.99$, and tidal deformability values are fixed to zeros.
The distance range is $10\,\mathrm{Mpc} \leq D_{\mathrm{L}} \leq 1000\,\mathrm{Mpc}$, yielding the median network optimal \ac{snr} value of $30$.
Only the luminosity distance parameter is marginalized over with the look-up table method since analytical phase marginalization is not applicable when gravitational-wave higher multipole moments are present.
The median sampling time is $198\,\mathrm{min}$.

Figure \ref{fig:pp_bbh} presents $P$-$P$ plots of all the source parameters. 
The $p$-values in the legend imply the observed deviations from the diagonal line is consistent with statistical errors, and the combined $p$-value is $0.42$.
This demonstrates our rapid parameter estimation framework can also be applied to full parameter estimation analysis of \ac{bbh} signal.

\section{Conclusion} \label{sec:conclusion}

In this paper, we have presented a rapid parameter estimation framework using multiple \ac{roq} bases of state-of-art GW signal models describing a broad range of CBC sources. Each basis is constructed over narrow parameter space to produce a significant speedup, while the union of the bases covers broad parameter space. Hence our framework can accelerate parameter estimation significantly without sacrificing the accuracy. Based on this idea, we have developed sets of \ac{roq} bases constructed over narrow chirp mass sub-domains. As demonstrated in Section \ref{sec:application}, our framework and new \ac{roq} bases enable improved source localization of \ac{bns} signal with the timescale of minutes, as well as more detailed parameter estimation taking into account general spin configurations, \ac{gw} higher multipole moments, and tidal deformation of colliding objects, with the timescale of hours. The combined use cases greatly improves the scalability of parameter estimation workflows during observing runs, especially as event rates increase due to improved detector sensitivity. Our multiple-\ac{roq}-bases framework has been implemented in one of the \ac{lvk} parameter estimation engines, \bilby{}, and that framework as well as our newly developed \ac{roq} bases are being employed by the automated parameter estimation analysis of the \ac{lvk} \ac{o4} alert system, circulating source location estimates to follow-up observers \cite{2023GCN.33816....1L, 2023GCN.33891....1L, 2023GCN.33919....1L, 2023GCN.34087....1L}.

One possible extension of this work is to extend the lower mass end of \phenomxphm{} bases. While constructing \phenomxphm{} bases in the lower mass region is computationally costly and consumes excessive amounts of memory, they are useful for rapid source localization of \ac{nsbh} signal taking into account orbital precession and \ac{gw} higher multipole moments. As we did for the other bases, directly constructing multi-banded bases can significantly reduce the memory consumption. Dimensionality reduction by a sophisticated choice of base waveforms, as done for the construction of the \phenomp{} and \phenompnrtidal{} bases, may also be possible. 
 
Another direction is to construct bases for other \ac{bns} waveform models, such as \texttt{SEOBNRv4ROM$\_$NRTidalv2} \cite{Bohe:2016gbl, Dietrich:2019kaq}, \texttt{IMRPhenomXP$\_$NRTidalv2}, and \texttt{SEOBNRv4T$\_$Surrogate} \cite{Lackey:2018zvw}. Those bases enable us to infer tidal deformability parameters with multiple waveform models, mitigating waveform systematics to obtain accurate constraints on the nuclear equation of state.

\acknowledgements{
We are grateful to Carl-Johan Haster for careful reading of the manuscript and useful feedbacks. SM and RS would like to thank the Institute for Computational and Experimental Research in Mathematics (ICERM) at Brown University for the support during their Fall 2020 Reunion Event, where considerable progress was made on this work. SM is supported by JSPS Grant-in-Aid for Transformative Research Areas (A) No.~23H04891 and No.~23H04893. This research was supported in part by the Australian Research Council Centre of Excellence for Gravitational Wave Discovery (OzGrav), through project number CE170100004. SS is a recipient of an ARC Discovery Early Career Research Award (DE220100241). CT is supported by an MIT Kavli Fellowship. AZ is supported by NSF Grant PHY-2308833. This work has been assigned preprint numbers LIGO-P2300205 and UTWI-27-2023. The authors are grateful for computational resources provided by the LIGO Laboratory and the Leonard E Parker Center for Gravitation, Cosmology and Astrophysics at the University of Wisconsin-Milwaukee supported by National Science Foundation Grants PHY-1626190, PHY-1700765, PHY-0757058 and PHY-0823459. This research has made use of data or software obtained from the Gravitational Wave Open Science Center (gwosc.org), a service of the LIGO Scientific Collaboration, the Virgo Collaboration, and KAGRA. This material is based upon work supported by NSF's LIGO Laboratory which is a major facility fully funded by the National Science Foundation, as well as the Science and Technology Facilities Council (STFC) of the United Kingdom, the Max-Planck-Society (MPS), and the State of Niedersachsen/Germany for support of the construction of Advanced LIGO and construction and operation of the GEO600 detector. Additional support for Advanced LIGO was provided by the Australian Research Council. Virgo is funded, through the European Gravitational Observatory (EGO), by the French Centre National de Recherche Scientifique (CNRS), the Italian Istituto Nazionale di Fisica Nucleare (INFN) and the Dutch Nikhef, with contributions by institutions from Belgium, Germany, Greece, Hungary, Ireland, Japan, Monaco, Poland, Portugal, Spain. KAGRA is supported by Ministry of Education, Culture, Sports, Science and Technology (MEXT), Japan Society for the Promotion of Science (JSPS) in Japan; National Research Foundation (NRF) and Ministry of Science and ICT (MSIT) in Korea; Academia Sinica (AS) and National Science and Technology Council (NSTC) in Taiwan.
}

\appendix

\section{Multi-band decomposition of ROQ likelihood ratio} \label{sec:multiband}

In this appendix, we obtain a multi-banded form of \ac{roq} likelihood ratio. In the multi-band approximation developed in \cite{Morisaki:2021ngj}, the total frequency range is divided into $B$ overlapping frequency bands $f^{(b)}_{\mathrm{s}} \leq f \leq f^{(b)}_{\mathrm{e}}~(b=1,2,\dots,B)$ with a set of smooth window functions $\{w^{(b)}(f)\}_{b=1}^{B}$. The $b$-th frequency band is constructed so that signal duration from the starting frequency $f^{(b)}_{\mathrm{s}}$ is smaller than a certain duration value $T^{(b)}$, where $T=T^{(1)}>T^{(2)}>\cdots>T^{(B)}$. The inner products, $(d_i, h_i(\theta))_i$ and $(h_i(\theta), h_i(\theta))_i$, are then approximated into the following forms (See Eqs. (24) and (45) of \cite{Morisaki:2021ngj}),
\begin{align}
&\begin{aligned}
&(d_i, h_i(\theta))_i \simeq \\
&\sum_{b=1}^{B} \frac{4}{T^{(b)}} \Re \left[\sum_{k=\ceil{f^{(b)}_{\mathrm{s}} T^{(b)}}}^{\floor{f^{(b)}_{\mathrm{e}} T^{(b)}}} w^{(b)}(f^{(b)}_k) \tilde{D}^{(b)\ast}_k h(f^{(b)}_k) \right],
\end{aligned} \label{eq:multiband_linear} \\
&\begin{aligned}
&(h_i(\theta), h_i(\theta))_i \simeq \\
&\sum^{B}_{b=1} \frac{4}{\hat{T}^{(b)}} \sum_{k=\ceil{f^{(b)}_{\mathrm{s}} \hat{T}^{(b)}}}^{\floor{f^{(b)}_{\mathrm{e}} \hat{T}^{(b)}}} w^{(b)}(\hat{f}^{(b)}_k) \tilde{I}_{\mathrm{c},k}^{(b)} \left|h(\hat{f}^{(b)}_k)\right|^2,
\end{aligned} \label{eq:multiband_quadratic}
\end{align}
where $f^{(b)}_k = k / T^{(b)}$, $\tilde{D}^{(b)\ast}_k$ is a quantity dependent on data and \ac{psd}, $\hat{T}^{(b)} = \min\left[2 T^{(b)}, T\right]$, $\hat{f}^{(b)}_k = k / \hat{T}^{(b)}$, and $\tilde{I}_{\mathrm{c},k}^{(b)}$ is a quantity dependent on \ac{psd}.

Substituting Eqs. \eqref{eq:rom_linear} and \eqref{eq:rom_quadratic} into Eqs. \eqref{eq:multiband_linear} and \eqref{eq:multiband_quadratic}, and substituting the approximate inner products into the log-likelihood-ratio \eqref{eq:log_likelihood_ratio}, we obtain the 
\begin{align}
&\ln \Lambda^{\mathrm{MB}}_{\text{ROQ}} =  \sum_{i=1}^{N_\mathrm{det}} \left[ L^{\mathrm{MB}}_i(\theta) - \frac{1}{2}Q^{\mathrm{MB}}_i(\theta) \right], \\
&L^{\mathrm{MB}}_i(\theta) =  \Re \left[\sum_{I=1}^{N_L}h_i(F_I;\theta)\,\omega^{\mathrm{MB}}_{I,i}(t_c)\right], \\
&Q^{\mathrm{MB}}_i(\theta) = \sum_{J=1}^{N_Q} |h_i(\mathcal{F}_J;\theta)|^2 \psi^{\mathrm{MB}}_{J,i},
\end{align}
where
\begin{align}
&\begin{aligned}
&\omega^{\mathrm{MB}}_{I,i}(t_{\mathrm{c}}) = \sum_{b=1}^{B} \frac{4}{T^{(b)}} \times \\
&\Re \left[ \sum_{k=\ceil{f^{(b)}_{\mathrm{s}} T^{(b)}}}^{\floor{f^{(b)}_{\mathrm{e}} T^{(b)}}} w^{(b)}\left(f^{(b)}_k\right) \tilde{D}^{(b) \ast}_k B_I\left(f^{(b)}_k\right) e^{-2\pi i f^{(b)}_k t_c} \right],
\end{aligned} \\
&\begin{aligned}
&\psi^{\mathrm{MB}}_I = \\
&\sum^{B}_{b=1} \frac{4}{\hat{T}^{(b)}} \sum_{k=\ceil{f^{(b)}_{\mathrm{s}} \hat{T}^{(b)}}}^{\floor{f^{(b)}_{\mathrm{e}} \hat{T}^{(b)}}} w^{(b)}\left(\hat{f}^{(b)}_k\right) \tilde{I}^{(b)}_{c,k} C_J(\hat{f}^{(b)}_k).
\end{aligned}
\end{align}
The integration weights, $\omega^{\mathrm{MB}}_{I,i}(t_{\mathrm{c}})$ and $\psi^{\mathrm{MB}}_{J,i}$, can be computed with the multi-banded \ac{roq} bases, $\{\{B_{I}(f^{(b)}_k)\}_k\}_{b=1}^{B}$ and $\{\{C_{J}(\hat{f}^{(b)}_k)\}_k\}_{b=1}^{B}$, and hence only those downsampled components need to be stored.

\section{Cusps in IMRPhenomPv2 waveforms} \label{sec:pv2_cusps}

\phenomp{} waveforms have cusps in a certain mass-spin space, where mass ratio $q$ is relatively low ($q \lesssim 0.4$) and total spin angular momentum projected onto the orbital angular momentum are negative. The cusps come from the Wigner coefficients $d^2_{2,m}(-\beta(f))$, where $\beta(f)$ is the opening angle between the total angular momentum and the orbital angular momentum at a \ac{gw} frequency $f$.

\begin{figure}[t]
	\centering
    \includegraphics[width=0.98\linewidth]{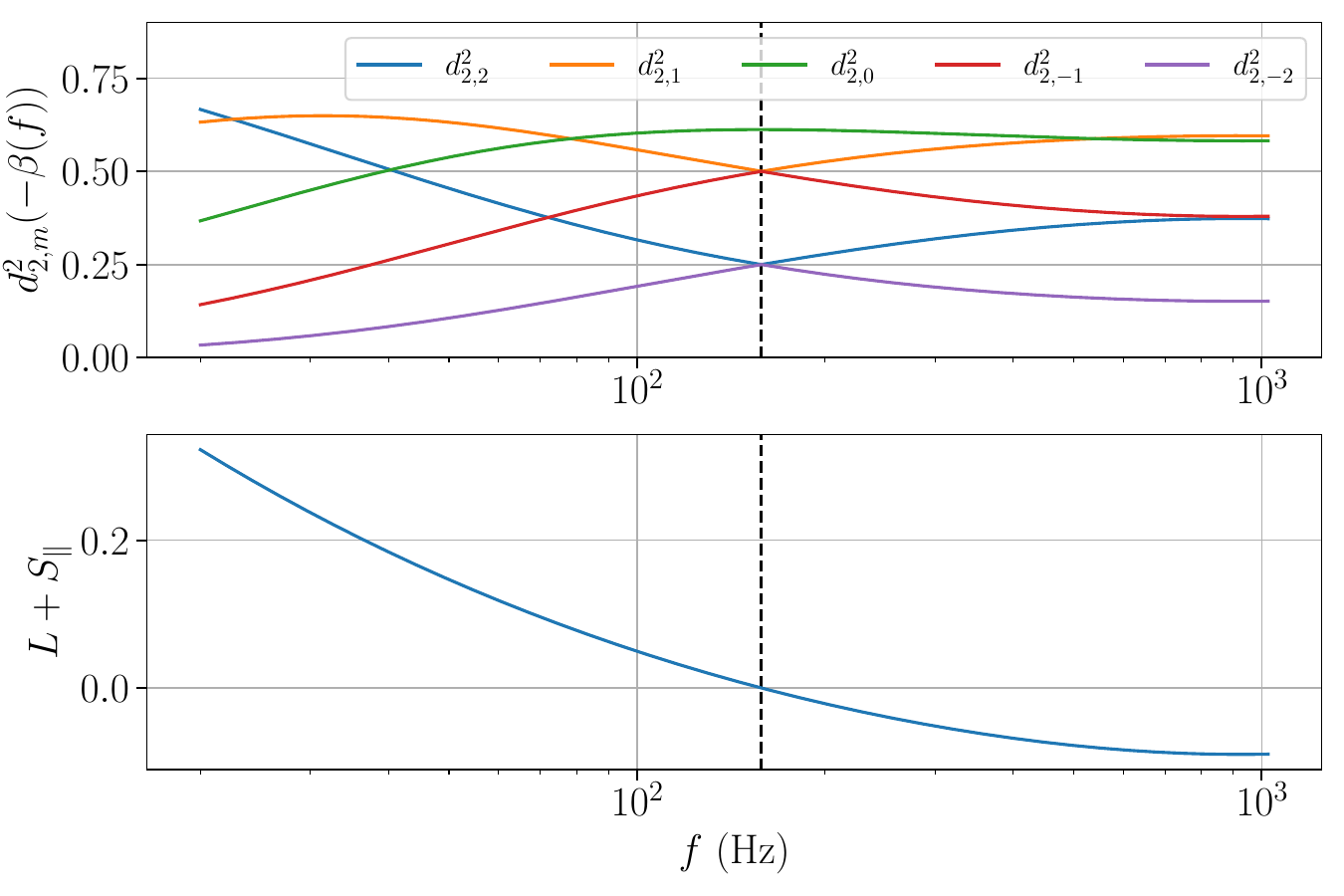}
	\caption{Wigner coefficients $d^2_{2,m}(-\beta(f))$ (top) and the total angular momentum projected onto the orbital angular momentum (bottom) of \phenomp{} for $m_1=8\Msun,\,m_2=1\Msun\,,\chi_1=-0.5\,,\chi_2=-0.5\,,\chi_{\mathrm{p}}=0.5$. The dashed vertical line indicates the frequency where the projected total angular momentum crosses $0$.} \label{fig:wigner}
\end{figure}

By the definition of $\beta(f)$, $\cos \beta(f)$ can be calculated as follows,
\begin{equation}
\cos \beta(f) = \pm \left(1 + \left(s(f)\right)^2\right)^{-\frac{1}{2}},
\end{equation}
where $s(f) = S_{\perp} / (L(f) + S_{\parallel})$, $L(f)$ is the norm of the orbital angular momentum at a \ac{gw} frequency $f$, and $S_{\parallel}$ and $S_{\perp}$ are the components of total spin angular momentum parallel with and orthogonal to the orbital angular momentum respectively. Mathematically the sign of $\cos \beta(f)$ should follow the sign of $L(f) + S_{\parallel}$. However in \phenomp{} the positive sign is always taken regardless of the sign of $L(f) + S_{\parallel}$. Thus $\cos \beta(f)$ has a cusp at a frequency where $L(f) + S_{\parallel}$ crosses $0$, and hence $d^2_{2,m}(-\beta(f))$ also has a cusp there since it depends on $\cos \beta(f)$, as shown in Fig. \ref{fig:wigner}.

In the parameter space we consider $L(f) + S_{\parallel}$ is always positive at $f=\flow=20\,\mathrm{Hz}$. Hence the necessary and sufficient condition for the existence of a waveform cusp is that the minimum of $L(f) + S_{\parallel}$ below $f=\fhigh$ is negative. In \phenomp{}, $L(f)$ is calculated with the non-spinning second-order Post-Newtonian formula,
\begin{equation}
\begin{aligned}
&L(f) = \frac{\eta (m_1 + m_2)^2}{v} \times \\
&~~ \left(1 + \left(\frac{3}{2} + \frac{\eta}{6}\right) v^2 + \left(\frac{27}{8} - \frac{19 \eta}{8} - \frac{\eta^2}{24}\right) v^4 \right),
\end{aligned}
\end{equation}
where $\eta = q / (1 + q)^2$ and $v = (\pi (m_1 + m_2) f)^{\frac{1}{3}}$. Within the frequency range from $\flow$ to $\fhigh$, it gets its minimum at $v=\hat{v}$, where
\begin{widetext}
\begin{equation}
\hat{v} = \min\left[ \sqrt{\frac{2 (9 + \eta - \sqrt{1539 - 1008 \eta - 17 \eta^2})}{3 (-81 + 57 \eta + \eta^2)}}, \left( \pi (m_1 + m_2) \fhigh\right)^{\frac{1}{3}}\right].
\end{equation}
\end{widetext}
On the other hand, $S_{\parallel} = m^2_1 \chi_1 + m^2_2 \chi_{2}$. Thus the mass-spin region where waveform has a cusp is expressed by
\begin{widetext}
\begin{equation}
\frac{\eta}{\hat{v}} \left(1 + \left(\frac{3}{2} + \frac{\eta}{6}\right) \hat{v}^2 + \left(\frac{27}{8} - \frac{19 \eta}{8} - \frac{\eta^2}{24}\right) \hat{v}^4 \right) + \frac{\chi_1 + q^2 \chi_2}{(1 + q)^2} < 0. \label{eq:pv2_exclude}
\end{equation}
\end{widetext}

\section{Base waveforms for IMRPhenomPv2}

\ac{gw} polarizations of \phenomp{} are given by
\begin{align}
h_+ (f) &= \frac{1}{2} \e^{-2 \iu \epsilon(f)} \left(T(f) + T_c(f)\right) h_{\mathrm{D}}(f), \\
h_{\times} (f) &= \frac{\iu}{2} \e^{-2 \iu \epsilon(f)} \left(T(f) - T_c(f)\right) h_{\mathrm{D}}(f),
\end{align}
where
\begin{align}
&\begin{aligned}
&T(f) = \\
&~~\sum_{m=-2}^{2} (-1)^m \e^{\iu m \alpha(f)} d^2_{2, m} (-\beta(f)) Y^2_{2,-m}(\theta_J, 0),
\end{aligned} \\
&T_c(f) = \sum_{m=-2}^{2} \e^{\iu m \alpha(f)} d^2_{2, m} (-\beta(f)) Y^2_{2,m} (\theta_J, 0),
\end{align}
and $Y^s_{l,m}(\theta, \phi)$ is spin-weighted spherical harmonics. Hence it is evident that $h_+ (f)$ and $h_{\times} (f)$ are linear combinations of $l_m(f)~(m=-2,-1,0,1,2)$ given by Eq. (\ref{eq:pv2_l}).

On the other hand, $\left|F_+ h_+ + F_{\times} h_{\times} \right|^2$ contains the following products,
\begin{align}
&\begin{aligned}
\left|h_{+/\times} (f)\right|^2 = \frac{1}{4} \bigg( &\left|T(f)\right|^2 + \left|T_c(f)\right|^2 \\
&~\pm 2 \Re \left[T^\ast(f) T_c(f)\right] \bigg) \left|h_{\mathrm{D}}(f)\right|^2,
\end{aligned} \\
&\Re \left[ h^\ast_+ (f) h_{\times} (f) \right] = \frac{1}{2} \Im \left[ T^\ast(f) T_c(f) \right] \left|h_{\mathrm{D}}(f)\right|^2.
\end{align}
$\left|T(f)\right|^2 + \left|T_c(f)\right|^2$ is linear combination of $q^{\mathrm{cos}}_{m,m'}(f)$ as shown below,
\begin{align}
&\left|T(f)\right|^2 + \left|T_c(f)\right|^2 \nonumber \\
&=\sum_{m,m'} \left[ (-1)^{m + m'} Y_{-m} Y_{-m'} + Y_m Y_{m'} \right] \nonumber \\
&\times d^2_{2,m}(-\beta(f)) d^2_{2,m'}(-\beta(f)) \e^{\iu (m - m') \alpha(f)} \nonumber \\
&= \sum_{m,m'} \left[ (-1)^{m + m'} Y_{-m} Y_{-m'} + Y_m Y_{m'} \right] \nonumber \\
&\times d^2_{2,m}(-\beta(f)) d^2_{2,m'}(-\beta(f)) \cos\left[(m - m') \alpha(f)\right] \nonumber \\
&\begin{aligned}
&=\frac{1}{2}\sum_{m,m'} \left[ (-1)^{m + m'} Y_{-m} Y_{-m'} + Y_m Y_{m'} \right] \\
&\times q^{\mathrm{cos}}_{m,m'}(f),
\end{aligned}
\end{align}
where $Y_m$ represents $Y^2_{2,m} (\theta_J, 0)$.
Similarly, $\Re \left[T^\ast(f) T_c(f)\right]$ is linear combination of $q^{\mathrm{cos}}_{m,m'}(f)$,
\begin{align}
&\Re \left[T^\ast(f) T_c(f)\right] \nonumber \\
&= \sum_{m,m'} (-1)^{m'} Y_m Y_{-m'} \nonumber \\
&\times d^2_{2,m}(-\beta(f)) d^2_{2,m'}(-\beta(f)) \cos \left[(m-m')\alpha(f)\right] \nonumber \\
&= \frac{1}{2} \sum_{m,m'} \left[ (-1)^{m'} Y_m Y_{-m'} + (-1)^{m} Y_{-m} Y_{m'} \right] \nonumber \\
&\times d^2_{2,m}(-\beta(f)) d^2_{2,m'}(-\beta(f)) \cos \left[(m-m')\alpha(f)\right] \nonumber \\
&\begin{aligned}
&=\frac{1}{4} \sum_{m,m'} \left[ (-1)^{m'} Y_m Y_{-m'} + (-1)^{m} Y_{-m} Y_{m'} \right] \\
&\times q^{\mathrm{cos}}_{m,m'}(f),
\end{aligned} \label{eq:Re_T_Tc_2}
\end{align}
and $\Im \left[T^\ast(f) T_c(f)\right]$ is linear combination of $q^{\mathrm{sin}}_{m,m'}(f)$,
\begin{align}
&\Im \left[T^\ast(f) T_c(f)\right] \nonumber \\
&= \sum_{m,m'} (-1)^{m'} Y_m Y_{-m'} \nonumber \\
&\times d^2_{2,m}(-\beta(f)) d^2_{2,m'}(-\beta(f)) \sin \left[(m-m')\alpha(f)\right] \nonumber \\
&= \frac{1}{2} \sum_{m,m'} \left[ (-1)^{m'} Y_m Y_{-m'} - (-1)^{m} Y_{-m} Y_{m'} \right] \nonumber \\
&\times d^2_{2,m}(-\beta(f)) d^2_{2,m'}(-\beta(f)) \sin \left[(m-m')\alpha(f)\right] \nonumber \\
&\begin{aligned}
&= \frac{1}{4}\sum_{m,m'} \left[ (-1)^{m'} Y_m Y_{-m'} - (-1)^{m} Y_{-m} Y_{m'} \right] \\
&\times q^{\mathrm{sin}}_{m,m'}(f).
\end{aligned} \label{eq:Im_T_Tc_2}
\end{align}
Hence $\left|F_+ h_+ + F_{\times} h_{\times} \right|^2$ is linear combination of $q^{\mathrm{cos}}_{m,m'}(f)$ and $q^{\mathrm{sin}}_{m,m'}(f)$.

\bibliographystyle{apsrev4-1}
\bibliography{reference}

\end{document}